\documentclass[epj,final]{svjour}
\usepackage{graphicx}
\usepackage[latin1]{inputenc}
\usepackage{amsmath}
\usepackage{amssymb}
\newcommand{\rmd}{{\rm d}}
\newcommand{\rmt}{{\rm t}}

\newcommand{\rmsol}{{\rm sol}}
\newcommand{\bfs}{{\bf s}}
\newcommand{\bfm}{{\bf m}}
\newcommand{\bfe}{{\bf e}}
\newcommand{\bfx}{{\bf x}}
\newcommand{\bfg}{{\bf g}}
\newcommand{\esol}{\bfe_{\rmsol}}

\newcommand{\bfA}{{\bf A}}
\newcommand{\bfB}{{\bf B}}
\newcommand{\bfC}{{\bf C}}
\newcommand{\bfD}{{\bf D}}

\newcommand{\bfM}{{\bf M}}
\newcommand{\bfU}{{\bf U}}
\newcommand{\bfW}{{\bf W}}
\newcommand{\bfV}{{\bf V}}
\newcommand{\bfR}{{\bf R}}

\newcommand{\bfUr}{\bfU^{\rm r}}
\newcommand{\bftUr}{^{\rmt}{\bfUr}}
\newcommand{\bfWr}{\bfW^{\rm r}}

\newcommand{\bftU}{{^{\rmt}\bfU}}
\newcommand{\bftV}{{^{\rmt}\bfV}}
\newcommand{\tM}{^{\rmt}\bfM}
\newcommand{\bfsr}{{\bfs^{\rm r}}}

\newcommand{\bftVr}{{^{\rmt}\bfV^{\rm r}}}
\newcommand{\bftVhit}{{^{\rmt}\bfV^{\rm hit}}}
\newcommand{\bfehit}{\bfe^{\rm hit}}
\newcommand{\bfshit}{\bfs^{\rm hit}}
\newcommand{\bfsrhit}{\bfsr^{\,\rm hit}}
\newcommand{\bfRhit}{\bfR^{\rm hit}}
\newcommand{\bfMhit}{\bfM^{\rm hit}}
\newcommand{\bfUhit}{\bfU^{\rm hit}}
\newcommand{\bftUhit}{{^{\rmt}\bfU^{\rm hit}}}
\newcommand{\bfWhit}{\bfW^{\rm hit}}

\begin{document}
\title{Matrix formalism and singular value decomposition for the location of 
gamma interactions in segmented HPGe detectors}
\titlerunning{Matrix formalism and singular value decomposition for the location of gamma interactions}
\author{ P. Désesquelles\inst{1} \and T.M.H. Ha\inst{1} \and K. Hauschild\inst{1} \and A. Korichi\inst{1}\and  F. Le Blanc\inst{2} \and A. Lopez-Martens\inst{1} \and A. Olariu\inst{2} \and C.M. Petrache\inst{2}}                     
\mail{Pierre.Desesquelles@in2p3.fr}%
\institute{CSNSM CNRS/IN2P3 and Université Paris Sud 11, 15 rue G. Clémenceau, 91405 Orsay, France \and IPNO CNRS/IN2P3 and Université Paris Sud 11, 15 rue G. Clémenceau, 91406 Orsay, France\\
\newline
On behalf of the AGATA Collaboration}
\date{Received: date / Revised version: date}
%
\abstract{
Modern coaxial and planar HPGe detectors allow a precise determination of the energies and  trajectories of the impinging gamma-rays. This entails the location of the gamma interactions inside the crystal from the shape of the delivered signals. This paper reviews the state of the art of the analysis of the HPGe response function and proposes methods that lead to optimum signal decomposition. The generic matrix method allows fast location of the interactions even when the induced signals strongly overlap.
} 
\maketitle

\section{Introduction}

Modern segmented germanium detectors have become an indispensable tool for both the measurement of gamma-rays energies \cite{Sim,Lee,Vet} and the 3D positioning of gamma-ray sources. Germanium detectors have evolved a lot since the sixties, notably by the increase of the crystal size and the possibility to cut the crystal into electrical segments. As for many physical detection devices \cite{Abb,Mar2,Ham,Cha}, the present improvements are mostly due to the development of on-line pulse shape analysis techniques. 

The segmentation of the germanium crystal is induced by electric contacts distributed on its surface. The volume of each resulting electrical segment is of the order of a several cubic centimeters.

Two main classes of such segmented germanium detectors exist. The first class, coaxial detectors, allow a precise determination of the energy (up to several MeV, for crystal length of the order of 10~cm) of the gamma-rays emitted by a nucleus and, in the case of a moving nucleus, their angle of emission so that measured energies may be properly Doppler corrected. In this frame, the detector array plays the role of a theodolite and a calorimeter. It must be able to detect in coincidence a large number of gamma-rays and to measure the characteristics of each photon individually. 

The arrays of planar or coplanar-segmented detectors play also the role of a positioning system of gamma sources. The ability of segmented germanium detectors to locate gamma-ray sources makes them valuable instruments for medical imaging, radioactive source search in the frame of national security or environmental monitoring. Planar array are typically made of layers of parallelepipedic segments. From the first two interactions of the incident gamma-ray, the Compton formula permits to determine a cone of possible directions to the source. The intersection of many such cones defines the position of the source \cite{Kha1,Tas}.

The location of the gamma-ray interactions (hits) inside the crystal is performed using pulse shape analysis. Indeed, Compton scattering, pair creation or photoelectric absorption generate a number of electron/hole pairs. The created charges migrate in the electric field to the contacts at the surface of the crystal. This motion induces a varying image charge on the electrodes belonging to the segment where the hit occurred and on the electrodes of the neighboring segments. The amplitudes of the signals are proportional to the deposited energy and the amplitude of the signals induced on the cathodes of the neighboring segments increases with the proximity of the hit. Hence, the location of the hit and the energy deposit can be deduced from the shape of the signals. When more than one hit occur simultaneously in the same segment or in neighboring segments, the total signal is the sum of the individual signals. Thus, the location of the interactions can be determined using a signal decomposition algorithm. The mathematical bases of such algorithms will be discussed in the following. Signal decomposition appears to be a more difficult problem in the case of coaxial detectors then in the case of planar detectors due to the variety of segment shapes and the complexity of the electrical field map. 

When the locations and the energy deposits have been determined for all hits, a so-called tracking code \cite{Sch2,Mar1,Wae} rebuilds the paths of the gamma-rays from hit to hit. Finally, the sum of the energy-deposits and the location of the first hits gives the characteristics of the gamma-rays.

This paper is organized in the following way. In the next section, we introduce the matrix formalism which allows to describe the mathematical link between signal shapes and locations of the hits. In the next section, the properties of the response function of the germanium detectors is deduced from the analysis of the transformation matrix. It will be shown that signal decomposition is an ill-posed problem. Moreover, signal decomposition entails the solving of a very large set of linear equations. The Singular Value Decomposition (SVD) method permits to solve both problems and to speed up the decomposition codes. In the forth section, we describe a complete protocol for on-line signal decomposition.

\section{Mathematical formalization}

\subsection{Introduction} 

\begin{figure}[htbp]
\begin{center}
\includegraphics[width=7.5cm]{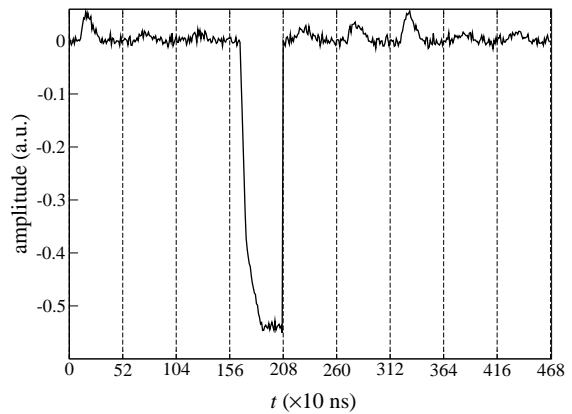}
\caption{Example of meta-signal. The fourth signal corresponds to the hit segment and the eight other signals correspond to its neighbors.}
\label{Fig meta-signal}
\end{center}
\end{figure}

Our goal is to determine the locations $(x_{i} , y_{i} , z_{i} )$ and the energy deposits $e_{i}$ of the $M$ interactions of the gamma-ray(s) inside the germanium crystal, knowing the sum of the $M$ individual signals. This signal decomposition is made possible thanks to two properties:  the amplitudes of the individual signals are proportional to the corresponding energy deposits, and the signals are additive, that is the signal actually delivered by the segments is the sum of the signals induced by each gamma interaction. A given hit induces signals in the segment where it occurred and in the neighboring segments. Thus it is useful to introduce the notion of meta-signal as simply the concatenation of the hit segment signal and its neighbor segment signals. An example is shown in fig. \ref{Fig meta-signal} where the signal of the hit segment runs from samples 157 to 208 and is concatenated with the signals of its eight neighbors. The number of samples (52) is chosen so that it includes the rise time of the hit segment signals, which maximum value is 37 samples, and at least 15 samples at the minimum of the signal as it gives the deposit energy. The number of neighboring segments is five for the first an the last layers of coaxial detectors.

The meta-signals will be denoted $s(t)$. The meta-signal can also be seen as a vector $\bfs$ whose components are simply the amplitude of the signal in the successive bins \footnote{In the following, vectors will be denoted by small bold letters and matrices by capital bold letters.}. This vector represents the whole information delivered by the detector. In the following, the meta-signal corresponding to a unit energy deposit at point $(x, y, z )$ will be noted $m(x, y, z , t)$.

\subsection{Linear system of equations}

\begin{figure}[htbp]
\begin{center}
\includegraphics[width=6.5cm]{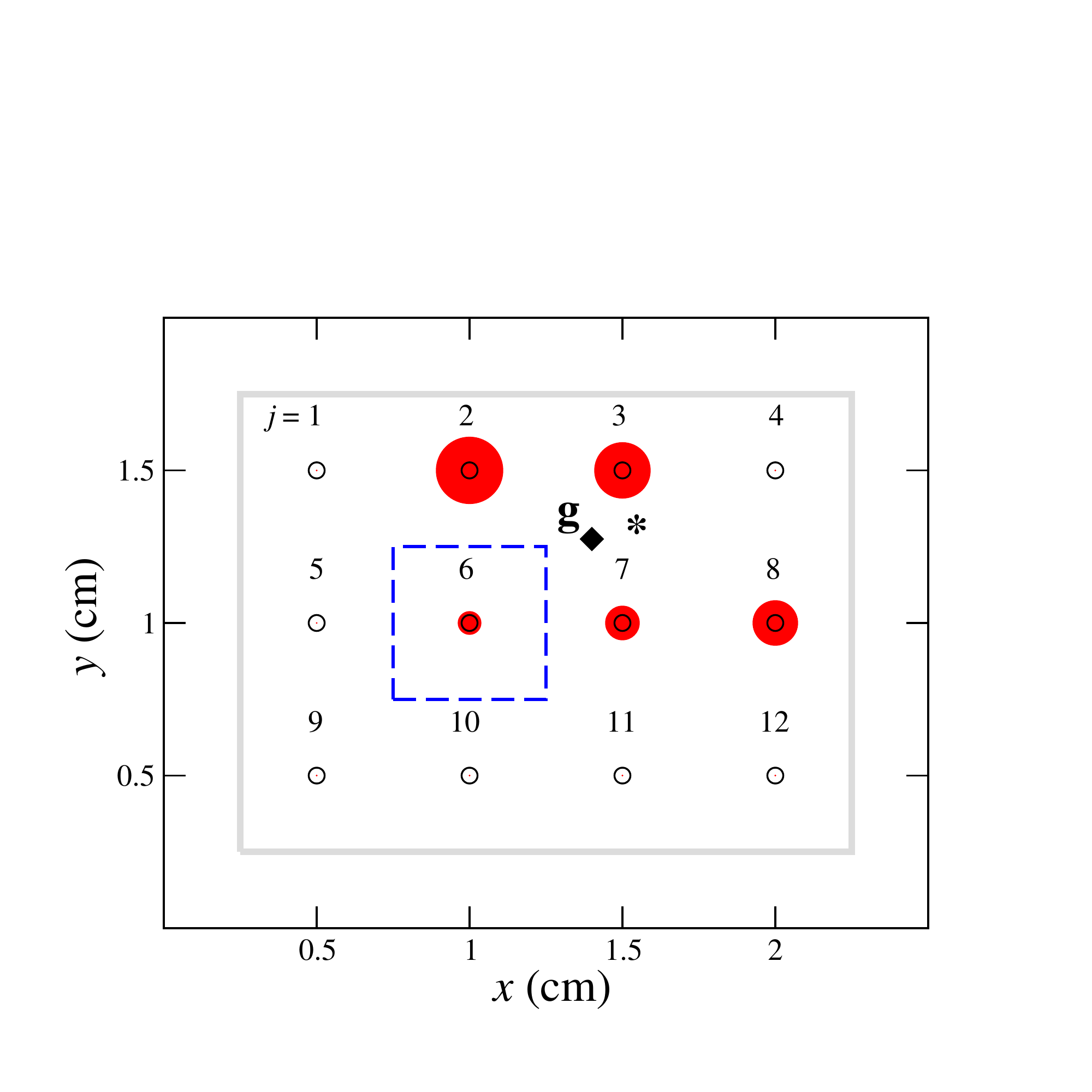}
\unitlength=1cm
\thicklines
\begin{picture}(5,0)
\put( 5.10,0.50){$\bfg$}
\put( 5.10,3.50){$\bfe$}
\put( 5.40,0.10){\line(0,1){1}}
\put( 5.40,1.60){\line(0,1){3.7}}
\put( 5.50,0.70){1.4}
\put( 5.50,0.30){1.275}
\put( 5.50,5.00){0}
\put( 5.50,4.70){.3}
\put( 5.50,4.40){.25}
\put( 5.50,4.10){0}
\put( 5.50,3.80){0}
\put( 5.50,3.50){.1}
\put( 5.50,3.20){.15}
\put( 5.50,2.90){.2}
\put( 5.50,2.60){0}
\put( 5.50,2.30){0}
\put( 5.50,2.00){0}
\put( 5.50,1.70){0}
\end{picture}
\caption{Example of calculation of the hit estimated location in a 2D rectangular segment. The open circles stand for the grid points. The dotted square around grid point 6 represents its voxel. The star indicates the actual location of the hit. The surfaces of the dots are proportional to the corresponding component in the $\bfe$ vector. The energy weighted barycenter (estimated location of the hit) is indicated by a diamond. The error of the method (resolution) is the distance between the diamond and the star.}
\label{Fig G}
\end{center}
\end{figure}
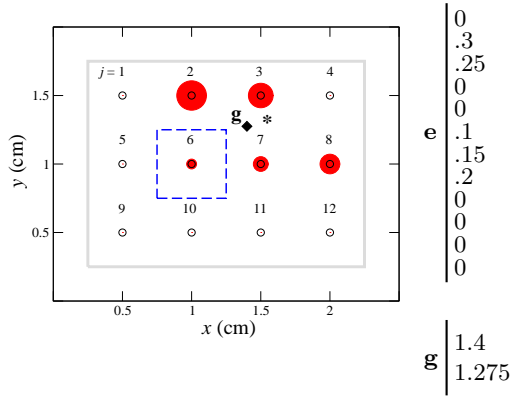

Using the linear and additive properties of meta-signals, one obtains: 

\begin{equation}
\label{s(t)}
 s(t) = \iiint e(x,y,z)\ m(x,y,z,t)\ \rmd x\, \rmd y\, \rmd z\ ,
\end{equation}

\noindent where $e(x,y,z)$ is the energy deposited at point $(x, y, z)$ and the sums run on the volume of the detector. If the meta-signals $m(x, y, z , t)$ are known, then the inversion of eq. (\ref{s(t)}) gives the energy deposits and their locations. The task of signal decomposition is to solve this so-called inverse problem. In actual applications, the meta-signals are calculated using a simulation code \cite{Med} or a crystal scanning system \cite{Bos,Cre2}. Thus, they are known only for discrete points $(x_{j} , y_{j} , z_{j})$ on the nodes of a given grid. The step of the grid is typically of a few millimeters. The previous equation becomes: 

\begin{equation}
\label{s(t) descrete}
 s(t) =\sum_{j} e_{j}\ m(j,t)\,,
\end{equation}

\noindent where $e_{j}$ is the energy deposited in the voxel surrounding the grid point $j$ as shown in fig. \ref{Fig G}. 

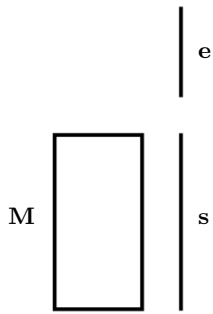
\begin{figure}[htbp]
\begin{center}
\includegraphics[width=7.5cm]{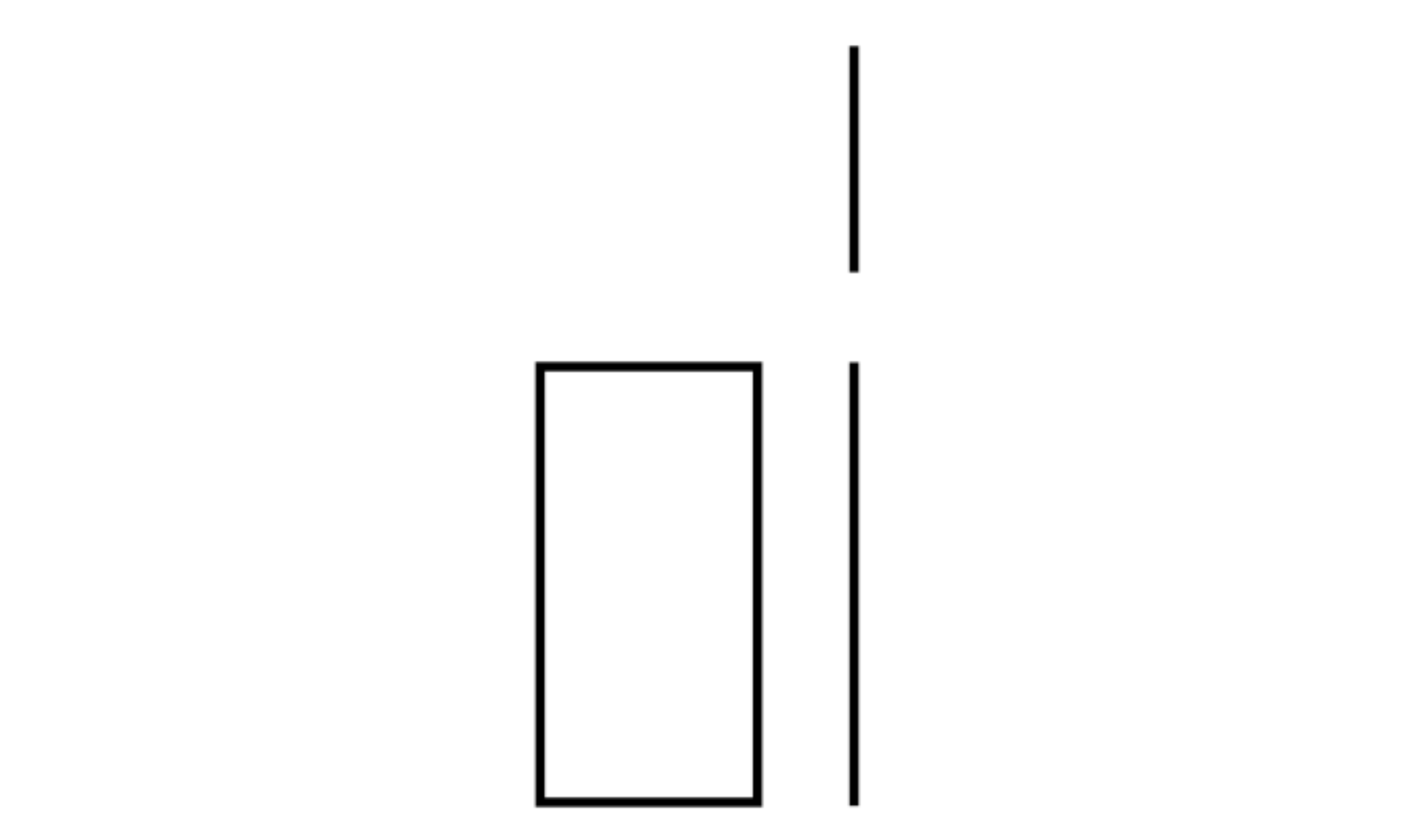}
\unitlength=1cm
\begin{picture}(5,0)
\put( 1.00,1.70){$\bfM$}
\put( 3.50,1.70){$\bfs$}
\put( 3.50,3.90){$\bfe$}
\end{picture}
\caption{Graphical sketch illustrating eq. (\ref{Me=s}). The bold box stands for the $\bfM$ matrix and the vertical segments for the $\bfe$ and $\bfs$ vectors. The logic of such diagrams is explained in Appendix A.}
\label{Fig Me}
\end{center}
\end{figure}

Noting $\bfm_{j}$ the basis meta-signal corresponding to a unit energy deposit at point $j$, this equation can be rewritten as:

\begin{equation}
\label{s(t) vector}
 \bfs = \sum_{j} e_{j} \ \bfm_{j}\,,
\end{equation}

\noindent and finally in a matrix form as:

\begin{equation}
\label{Me=s}
\bfM\, \bfe= \bfs\,,
\end{equation}

\noindent where $\bfM$ is the transformation matrix. The $j^{\rm th}$ column of $\bfM$ is $\bfm_{j}$ (fig. \ref{Fig Me}).

Signal decomposition consists in solving this matrix linear system, that is to find the components of the $\bfe$ vector. An example of energy vector is shown in fig. \ref{Fig G}. 

For this purpose, many algorithms have been developed, (grid search \cite{Bev,Ven}, matrix method \cite{Ola,Dox,Kha2}, wavelet decomposition \cite{Ger}, ...) some of them using artificial intelligence methods (genetic algorithms \cite{Kro}, neural network \cite{Sch1}, ...). In fact, artificial intelligence methods appeared to be efficient but slow and thus applicable only in the case when the interaction location is performed off-line. For real-time applications, the most used methods are improved forms of grid search. Unfortunately, these methods are not well adapted to the situations when more than one hit occurs in the same segment or in neighboring segments. 

In principal, the general solution of eq. (\ref{Me=s}) is obtained by inverting matrix $\bfM$ \cite{Ola}. Yet, the mere inversion method would require a very long computation time and would give an unphysical solution. In the following, we will see how these difficulties can be overcome. 

\subsection{Properties of the solution vector}

The components of the solution vector represent the energy deposit in each voxel of the grid. Of course, most of the $e_{j}$ components are equal to zero, and, as they represent energies, the non-null components are positive \footnote{The total energy deposited in a segment is directly deduced from the amplitude of the charge signal.}. In the hypothetical case when the hit occurs exactly on a grid point, only the corresponding solution component is different from zero and, as the basis signals correspond to unit energies, this component is equal to the energy deposit. In the real case, the hit location does not belong to the grid. The strategy of most algorithms, such as the grid search, is to find the signal from the grid that best matches the measured signal. The corresponding grid point is considered to be the closest to the actual gamma hit. In this case, the precision on the hit location is directly connected to the grid step. Using the matrix formalism, a more accurate location is possible. Indeed, after solving of eq. (\ref{Me=s}), more than one component of the solution vector can be different from zero, even for a single hit. The estimator of the hit location is then calculated as the energy weighted barycenter (fig. \ref{Fig G}) of the components: 

\begin{equation}
\label{G}
\bfg = \frac{\sum_{j} e_{j}\,\bfx_{j}}{\sum_{j} e_{j}}\,,
\end{equation}

\noindent where the $\bfx_{j}$ are the three dimensional locations of the grid points. The denominator is equal to the deposited energy. 

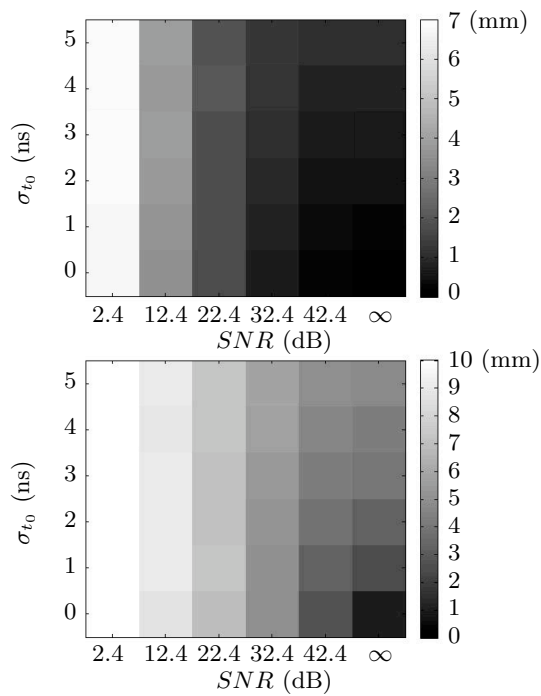
\begin{figure}[htbp]
\begin{center}
\includegraphics[width=6cm]{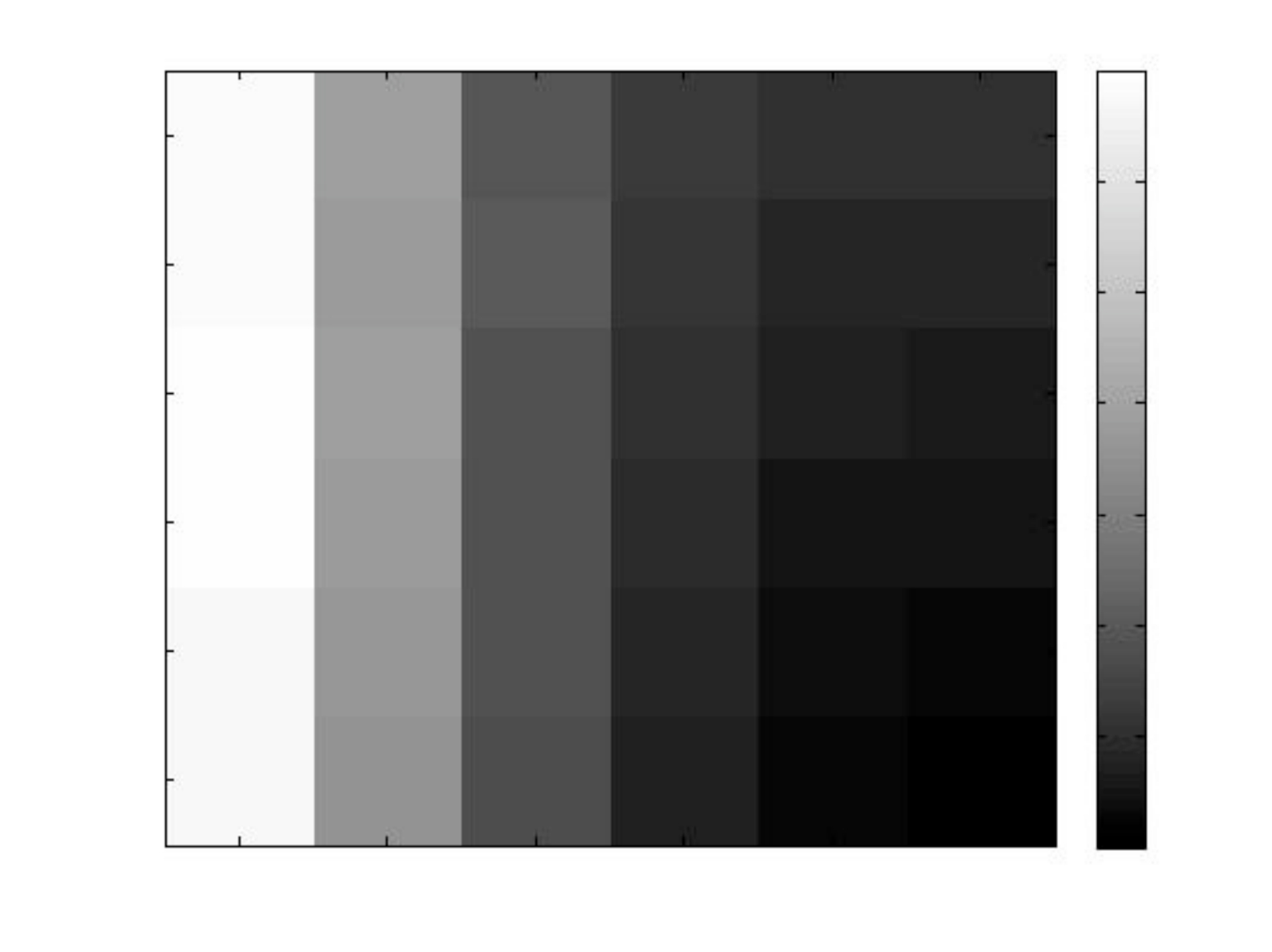}
\includegraphics[width=6cm]{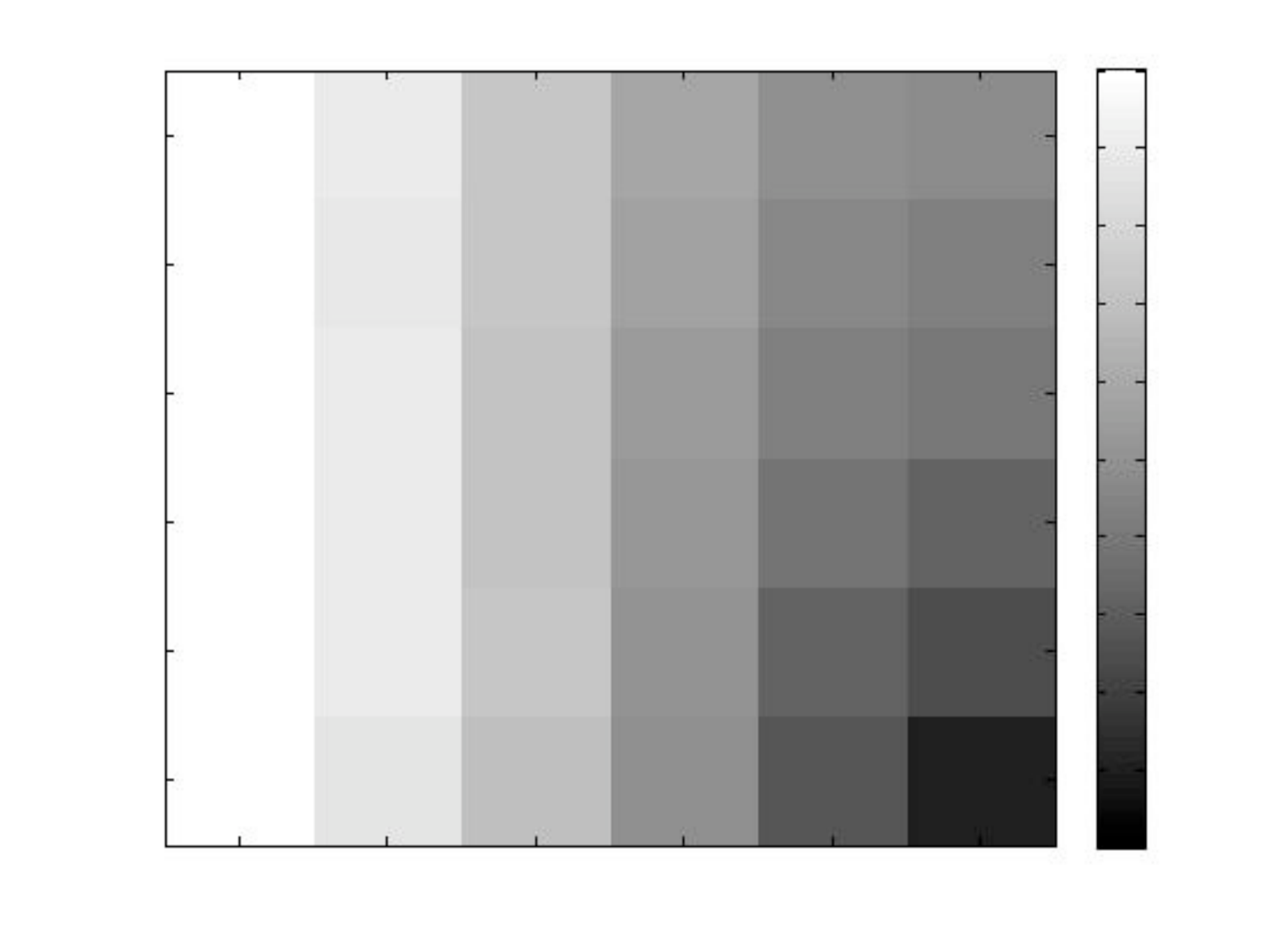}
\unitlength=1cm
\begin{picture}(5,0)
\put(  2.00,4.70){$SNR$ (dB)}
\put(  0.37,5.05){2.4}
\put(  1.05,5.05){12.4}
\put(  1.75,5.05){22.4}
\put(  2.45,5.05){32.4}
\put(  3.15,5.05){42.4}
\put(  4.00,5.05){$\infty$}
\put( -0.70,6.60){\rotatebox{90}{$\sigma_{t_{0}}$ (ns)}}
\put( -0.00,8.71){5}
\put( -0.00,8.10){4}
\put( -0.00,7.48){3}
\put( -0.00,6.87){2}
\put( -0.00,6.26){1}
\put( -0.00,5.65){0}
\put( 5.05,8.95){7 (mm)}
\put( 5.05,8.44){6}
\put( 5.05,7.91){5}
\put( 5.05,7.39){4}
\put( 5.05,6.87){3}
\put( 5.05,6.36){2}
\put( 5.05,5.85){1}
\put( 5.05,5.35){0}

\put(  2.00,0.20){$SNR$ (dB)}
\put(  0.37,0.55){2.4}
\put(  1.05,0.55){12.4}
\put(  1.75,0.55){22.4}
\put(  2.45,0.55){32.4}
\put(  3.15,0.55){42.4}
\put(  4.00,0.55){$\infty$}
\put( -0.70,2.10){\rotatebox{90}{$\sigma_{t_{0}}$ (ns)}}
\put( -0.00,4.17){5}
\put( -0.00,3.56){4}
\put( -0.00,2.94){3}
\put( -0.00,2.33){2}
\put( -0.00,1.72){1}
\put( -0.00,1.11){0}
\put( 5.05,4.45){10 (mm)}
\put( 5.05,4.09){9}
\put( 5.05,3.72){8}
\put( 5.05,3.35){7}
\put( 5.05,2.98){6}
\put( 5.05,2.62){5}
\put( 5.05,2.25){4}
\put( 5.05,1.89){3}
\put( 5.05,1.53){2}
\put( 5.05,1.16){1}
\put( 5.05,0.80){0}
\end{picture}
\caption{Resolution (top panel) and spread of the cloud of non-zero components of the $\bfe$ vector (bottom panel) as a function of the signal-to-noise ratio and of the standard deviation of the signal time jitter. These results are obtained by the NNLS solving of eq. (\ref{Me=s}) using MGS simulations of AGATA signals. Assuming a noise level of 3 keV sigma, the signal-to-noise bins correspond, from left to right, to about 10~keV, 33~keV, 100~keV, 333~keV and 1~MeV gamma hits.}
\label{Fig ed_rI_snr_t0}
\end{center}
\end{figure}

Of course, due to the uncertainties on the measured signal (electric noise, time alignment \cite{Sch1,Des4,Des6}, cross-talk \cite{Bru1}, etc.) and on the transformation matrix (grid discretization), it is not possible to find the exact position of hits. Both the resolution (i.e. the r.m.s. distance between the estimated and the actual hits) and the spread of the cloud of non-zero components of the solution vector, shown as dots in fig. \ref{Fig G}, depend on the alterations of the detected signal. As an example, fig.  \ref{Fig ed_rI_snr_t0} shows the evolution of the resolution and of the cloud spread as a function of the signal-to-noise ratio and of the time jitter of the signals. The time jitter applies to the meta signal as a whole. The time jitter between segments is much smaller and does not influence the resolution. The data are simulated using the MGS code. The deterioration of the resolution appears to be mainly due to the noise whereas the spread of the cloud is also sensitive to the amplitude of the time jitter. Large cloud spreads do not allow the separation of hits, which lie close together, as will be seen in section 4.

In the following, we focus on the mathematical properties of eq. (\ref{Me=s}) and on its optimum solving method.

\subsection{Least square solving}

The number of unknowns in eq.  (\ref{Me=s}) (columns in the $\bfM$ matrix) is the number of grid points and the number of equations (lines in the $\bfM$ matrix) is the number of samples in the meta-signal. This consideration gives a first upper limit to the number of grid points that can be accommodated so that the linear system is not underdetermined. Hence, the $\bfM$ matrix for a segment can be made vertical rectangular.  Due to the signal alterations, eq. (\ref{Me=s}) has no exact solution, thus, we are searching for the solution $\esol$ that minimizes the residue $\|\bfM\,\esol - \bfs\|$.

If there were no constraints on the components of the solution vector then the least square solution would be given by: 

\begin{equation}
\label{esol}
\esol = (\tM\,\bfM)^{-1}\ \tM\ \bfs\,,
\end{equation}

\noindent  where $\tM$ is the transpose of matrix $\bfM$. In fact, this algebraic solving is not valid since, due to the noise and the uncertainties that affect the system, the resulting solution would most probably have negative components. Different algorithms are meant to find non-negative solutions to this kind of linear system (such as backtracing \cite{Des2}) but most of them are too slow (the signal decomposition must be realized on-line, that is in a few milliseconds). To our knowledge, the best compromise algorithms are NNLS (nonnegative least squares \cite{Law}) and NNLC (nonnegative least chi-square \cite{Des5}), as they maximize the number of null components and minimize the size of the matrices to be inverted and thus the computing time. 

\subsection{Uniqueness of the solution}

\begin{figure}[htbp]
\begin{center}
\includegraphics[width=7.5cm]{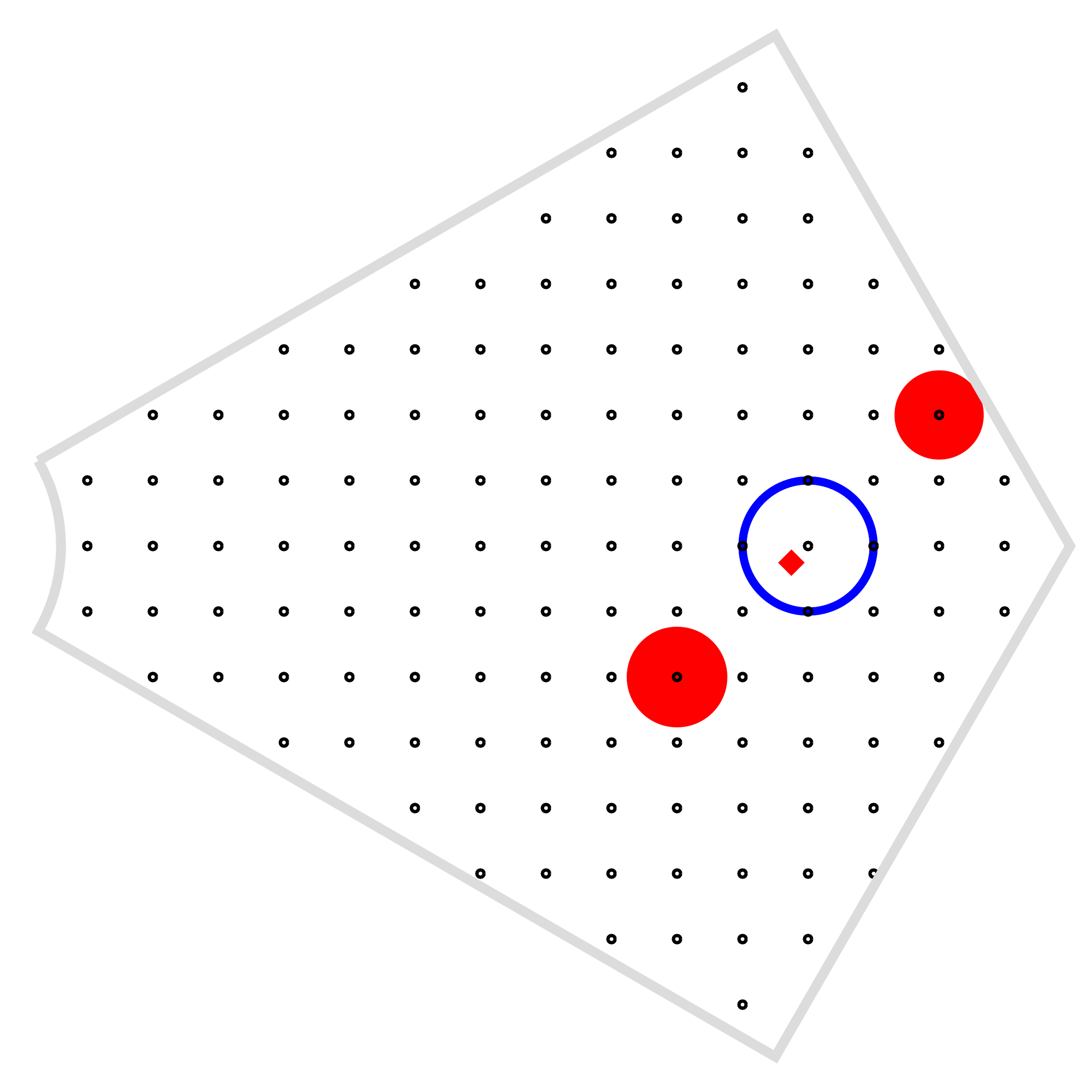}
\includegraphics[width=7.5cm]{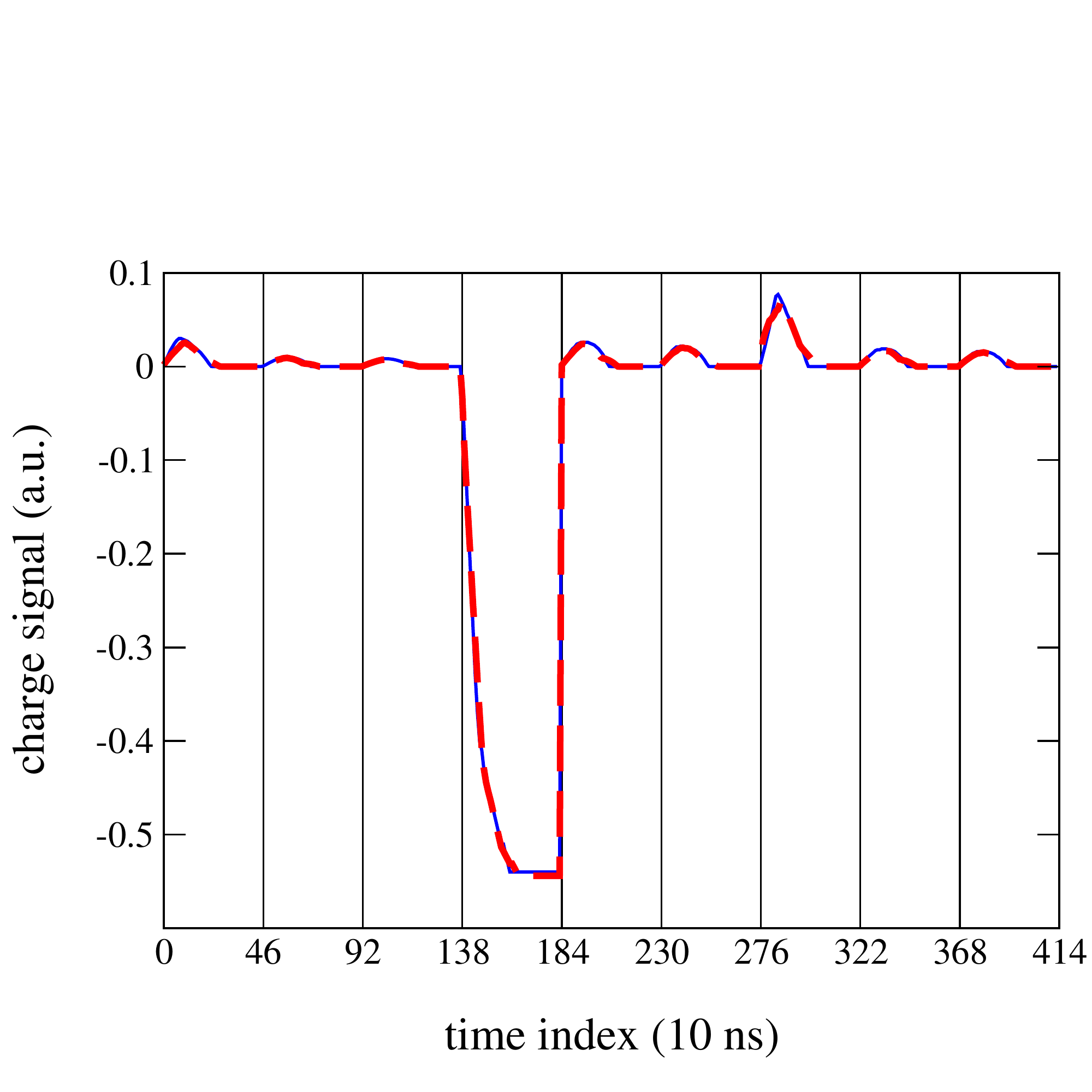}
\caption{Example of two hits giving the same signal as a single hit in the low sensitivity zone of a coaxial detector segment. The upper figure shows the location of the grid points (small open dots), of the single hit (open blue disk) and the two hits (red disks). The surfaces of the disks are proportional to the energy deposits. The barycenter of the two hits is indicated by a diamond. The resulting almost over-imposed meta-signals  are shown in the lower figure, with the same colors (dotted red line: two hits, thin blue line: one hit). These signals were generated using the MGS \cite{Med} code (neither the noise nor the electronic response function are taken into account here).}
\label{Fig 2 solutions}
\end{center}
\end{figure}

We will now discuss the characteristics of the matrix $\bfM$ that entail the uniqueness and the stability of the solution. The properties of $\bfM$ are induced both by the response function of the segments and by the choice of the number and the locations of the grid points. When the response function is not bijective then, whatever the algorithm and grid choices, the hits cannot be unambiguously located. This situation arises in some particular cases in coaxial and large segment planar detectors: the signal resulting from the addition of two 
hits may be very similar to the signal resulting from a single interaction at the barycenter of the two hits (see fig. \ref{Fig 2 solutions}). This explains why the determination of the number of hits within one segment is often difficult and some times impossible. If one, however, is searching for single hit events\footnote{Events are defined as sets of hits induced by gamma-rays entering simultaneously the crystal.}, then, whatever the found solution, the barycenter of the components is approximately the same (fig. \ref{Fig 2 solutions}).

The fact that the response function does not have a one to one relation between resulting pulse shape and grid locations, is equivalent to the fact that, whatever the chosen grid size, the matrix $\bfM$ is ill-conditioned. This does not mean that more than one solution is the usual situation. Indeed, on the one hand, the number of solutions is reduced by taking into account the physical constraints: the components have to be non-negative and most of them are equal to zero. On the other hand, indiscernability occurs only in the low sensitivity zones of the detector.

\section{Solution using singular value decomposition}
\label{Sec SVD}

\subsection{SVD principles}

We have seen that the $\bfM$  matrix has two main defects: on the one hand, it is large, thus the calculation of its inverse is irretrievably long and, on the other hand, small fluctuations on the detected signal induce large uncertainties on the hit locations. Both of these problems can be corrected using singular value decomposition. In the following, we will see how this technique can be adapted to our purpose. 

\begin{figure}[htbp]
\begin{center}
\includegraphics[width=7.5cm]{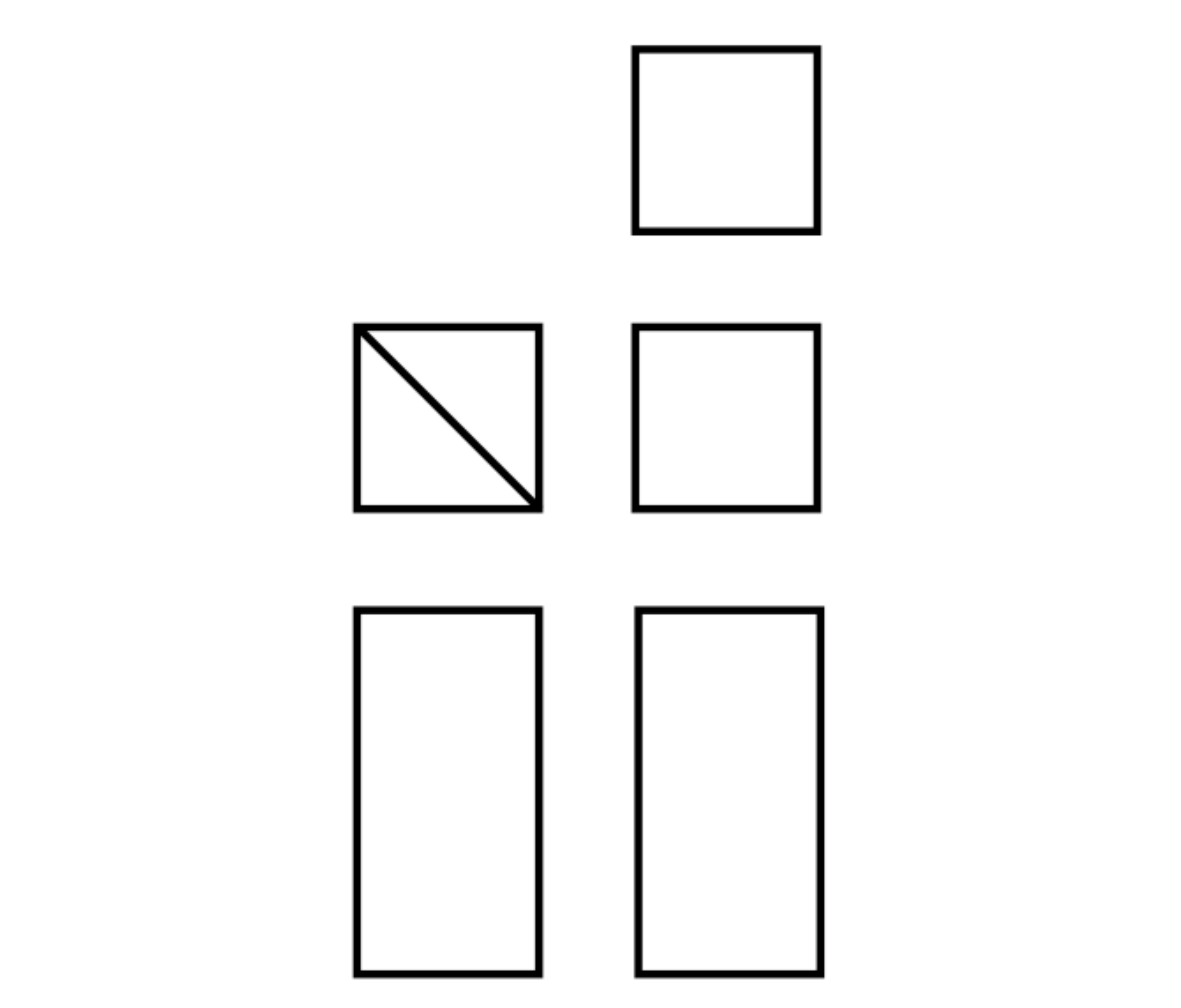}
\unitlength=1cm
\begin{picture}(5,0)
\put( 4.20,1.70){$\bfM$}
\put( 0.40,1.70){$\bfU$}
\put( 0.40,4.00){$\bfW$}
\put( 4.20,4.00){$\bfW\ \bftV$}
\put( 2.20,5.80){$\bftV$}
\put( 1.30,3.80){$\bf 0$}
\put( 1.80,4.25){$\bf 0$}
\end{picture}
\caption{Graphical sketch of the singular value decomposition in the case of a vertical $\bfM$ matrix (eq. (\ref{SVD})). The diagonal line in the $\bfW$ matrix symbolizes the set of singular values.}
\label{Fig svd}
\end{center}
\end{figure}
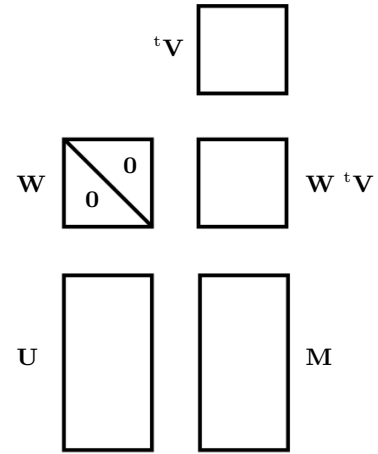

Any square or rectangular matrix can be decomposed, as shown in fig. \ref{Fig svd}, into the product of three matrices: 

\begin{equation}
\label{SVD}
\bfM = \bfU\, \bfW\, \bftV\,,
\end{equation}

\noindent such that $\bfW$ is a diagonal positive matrix whose diagonal components are arranged in a decreasing order, and $\bfU$ and $\bfV$ are column-orthonormal matrices ($\bftU\bfU=\bftV\bfV=\bf 1$). 

The set of singular values $w_{i}$, that is the values of the diagonal matrix, is unique. Singular value decomposition has many applications in Physics ranging from the prediction of the perturbation growth \cite{Far}, to Principal Component Analysis \cite{Des1,Des8} (inertia analysis in multivariate space) or genomic analysis \cite{Alt}. 

Once the matrix is decomposed, the least square solution vector could be very conveniently calculated as:

\begin{equation}
\label{e=VW-1Us}
\esol = \bfV\,\bfW^{-1}\,\bftU\,\bfs\,,
\end{equation}

\noindent where $\bfW^{-1}$ is the diagonal matrix of the $1/w_{i}$ (when a singular value is equal to zero, the corresponding component in the inverse matrix is also zero). Unfortunately, this solution does not necessarily respect the physical constraints proper to our pulse shape analysis problem. However, as we will see in the following, SVD remains very useful for signal decomposition. 

The first advantage of SDV is that it allows to transform the rectangular system of eq. (\ref{Me=s}) into a square one and to lower the dimension of the matrix to be inverted in the case when the number of samples of the meta-signals is greater than the number of grid points in the segment. Indeed, from eqs. (\ref{Me=s}), and (\ref{SVD}), one obtains:
\begin{eqnarray}
\label{tVe=W-1tUs}
&\bftV\,\bfe &=\bfW^{-1}\, \bftU\,\bfs\,.
\end{eqnarray}

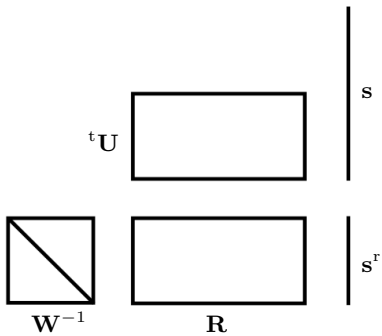
\begin{figure}[htbp]
\begin{center}
\includegraphics[width=7.5cm]{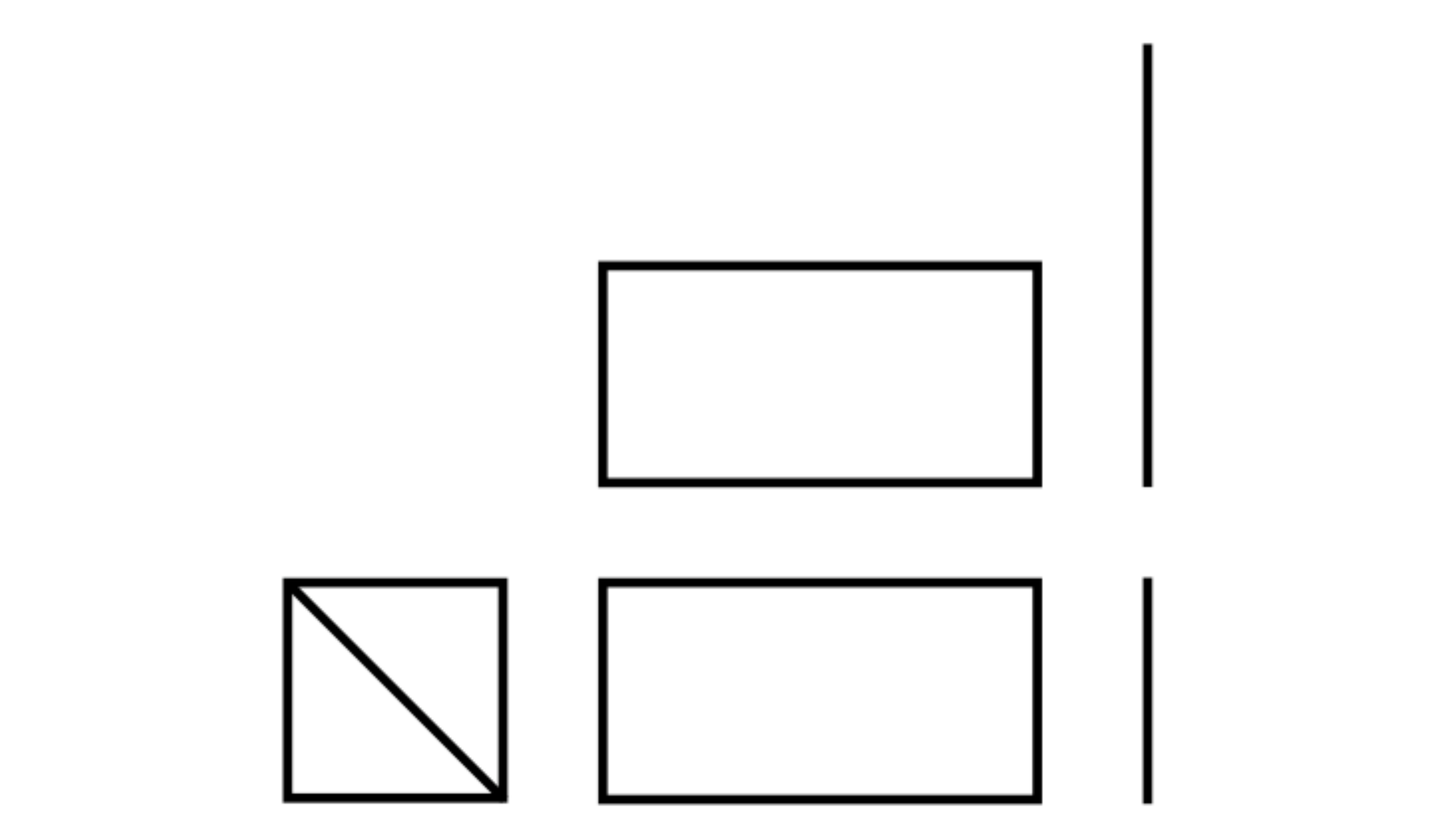}
\unitlength=1cm
\begin{picture}(5,0)
\put( 1.30,2.60){$\bftU$}
\put( 0.55,0.20){$\bfW^{-1}$}
\put( 2.85,0.20){$\bfR$}
\put( 4.90,3.30){$\bfs$}
\put( 4.90,1.00){$\bfsr$}
\end{picture}
\caption{Reduction of the size of the signal with matrix $\bfR$.}
\label{Fig Rs}
\end{center}
\end{figure}

The matrix $\bfR =\bfW^{-1}\, \bftU$ has the effect of reducing the size of the signal, as shown in fig. \ref{Fig Rs}, in an optimum way, i.e. keeping the whole relevant signal-information. The reduced signal will be denotes $\bfsr$.

Finally, a new smaller linear system can be substituted to eq. (\ref{Me=s}): 
\begin{eqnarray}
\label{tVe=sr}
&\bftV\,\bfe &=\bfsr\,.
\end{eqnarray}

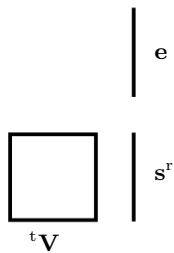
\begin{figure}[htbp]
\begin{center}
\includegraphics[width=7.5cm]{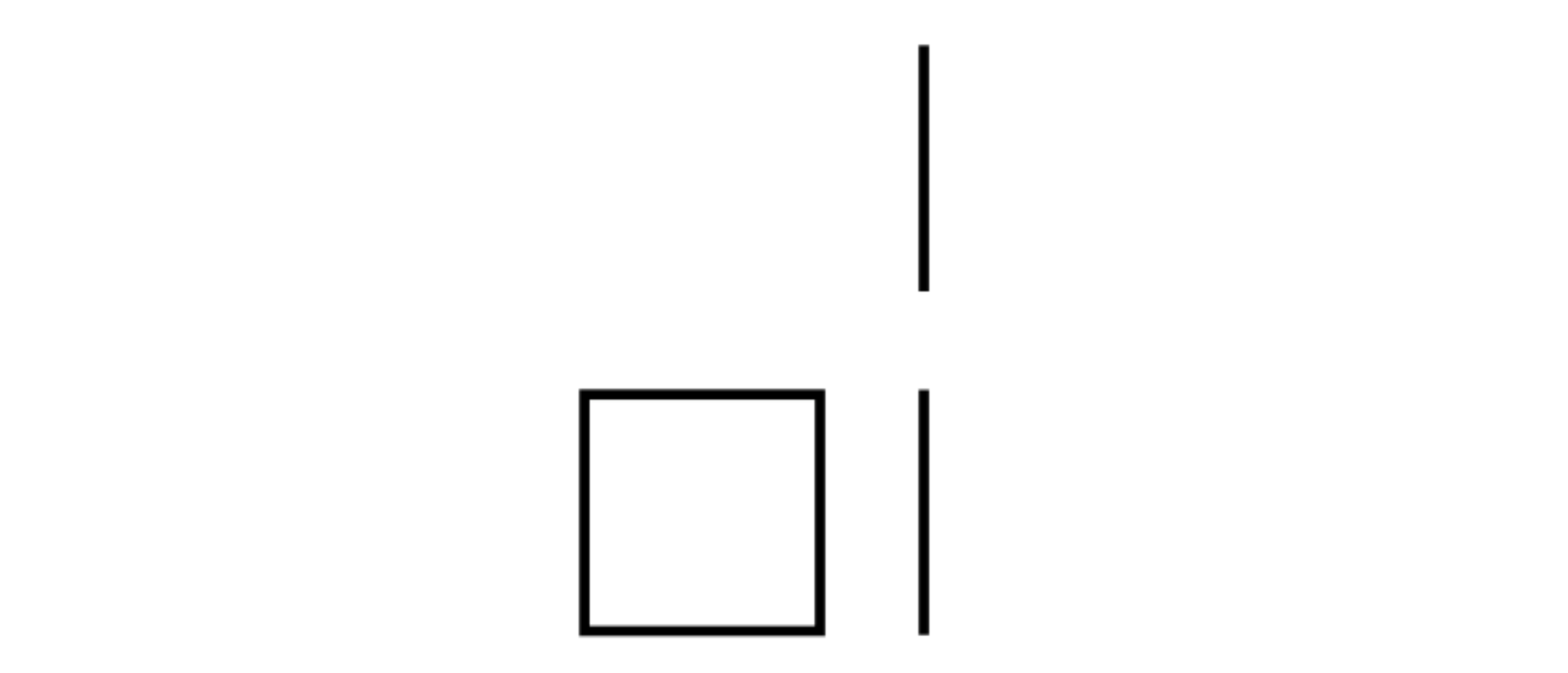}
\unitlength=1cm
\begin{picture}(5,0)
\put( 3.45,2.70){$\bfe$}
\put( 1.80,0.15){$\bftV$}
\put( 3.45,1.10){$\bfsr$}
\end{picture}
\caption{Graphical sketch of eq. (\ref{tVe=sr}). The size of the linear system is reduced with respect to fig. \ref{Fig Me}.}
\label{Fig tVe}
\end{center}
\end{figure}

The number of lines in the system is decreased as can be seen by comparing fig. \ref{Fig tVe} to fig. \ref{Fig Me}. Each column of $\bftV$ is the reduced signal corresponding to a unit energy deposit on a given grid point. This matrix plays the same role for reduced signals as $\bfM$ for signals.

In the following, we will see that the size of the system can be reduced even more using the so-called SVD truncation. It will be shown that the maximum reduction is connected to the condition number of the matrix. 

\subsection{Properties of the germanium response function}

An important characteristic of linear systems is the way they transmit the uncertainties, such as the noise, present on the right-hand term $\bfs$, to the solution $\bfe$. The maximum amplification coefficient for the relative uncertainties is called the condition number:

\begin{equation}
\label{C}
C = {\rm max_{d\bfs}}\left(\frac{\frac{\|\bfM^{-1}\rmd\bfs\|}{\|\bfM^{-1}\bfs\|}}{\frac{\|\rmd \bfs\|}{\|\bfs\|}}\right).
\end{equation}

\noindent which is the ratio of the relative fluctuations on the solution (the fluctuations are measured as the norm of the vector)  and on the signal. As can be seen in eq. (\ref{e=VW-1Us}), the amplification increases with the inverse of the singular values. In fact, it can be shown that the condition number is equal to the ratio between the largest and the lowest non-zero singular values. In the case of germanium detectors, the condition numbers may actually reach very high values. Therefore, a prior mathematical treatment of the transformation matrix is necessary to ensure the reliability of the solution. Singular Value Decomposition permits such a treatment.

An optimum way to lower the condition number, introducing a minimum bias on the solution, is simply to set to zero the lowest singular values. The condition number becomes: 

\begin{equation}
\label{C=w1/wr}
C = \frac{w_{1}}{w_{r}}\,,
\end{equation}

\noindent where $r$ is the index of the lowest non-zero singular value. 

Two types of ill-conditioned problems exist, which have to be addressed with different solving methods. The first includes the rank deficient problems which are characterized by a transformation matrix having two, well separated, groups of large and small singular values. In this case, the numerical rank is equal to the number of large singular values (and $r$ is fixed to this value). The second one includes the discrete ill-posed problems, for which the set of singular values decreases smoothly.  

\begin{figure}[htbp]
\begin{center}
\includegraphics[width=7.5cm]{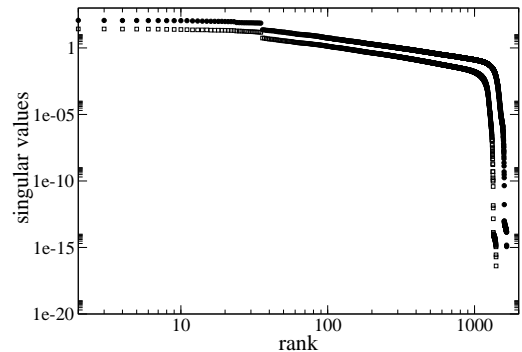}
\caption{Typical set of singular values for a 36 segment germanium coaxial crystal (boxes: 5~mm cubic grid, dots: 2~mm cubic grid). The signal are generated using the MGS code.}
\label{Fig singval_M}
\end{center}
\end{figure}

In fig. \ref{Fig singval_M}, we show the singular values for a 36 segment coaxial germanium detector for 2 mm and 5 mm cubic grids (here, the $\bfM$ matrices have respectively 41874 and 2544 columns for 1872 lines). The detector signals are simulated using the MGS simulation. The two distributions being very close, within a constant factor, we can say that the 2~mm grid adds little information with respect to the 5~mm grid. The condition number of the matrix is of the order of $10^{16}$, which is very bad, thus SVD truncation is indispensable. The singular values distributions show the same drop after one thousand. Thus the value of the cut-off $r$ should not be larger than one thousand. In fact, due to uncertainties in the $\bfm_{j}$ signals, the cut-off has to be even lower, as will be shown in section 3.4. 

A clear gap appears between the first 36 singular values and the next one. It means that, from the signal shapes, it is very easy to know which segment was hit, and, as $w_{i}/w_{36}$ is close to one, this determination is robust with respect to the signal noise. This does not come as a surprise. 

As can be seen the transformation matrix of a coaxial detector cannot be characterized by a numerical rank. Germanium signal decomposition is an ill-posed problem. Hence, the value of $r$ will have  to be tuned in order to find the right balance between the precision of the solution vector $\esol$ and the amount of fluctuations on its components. 

We now turn to the implementation of this conditioning improvement method.

\subsection{SVD truncation}
 
\begin{figure}[htbp]
\begin{center}
\includegraphics[width=7.5cm]{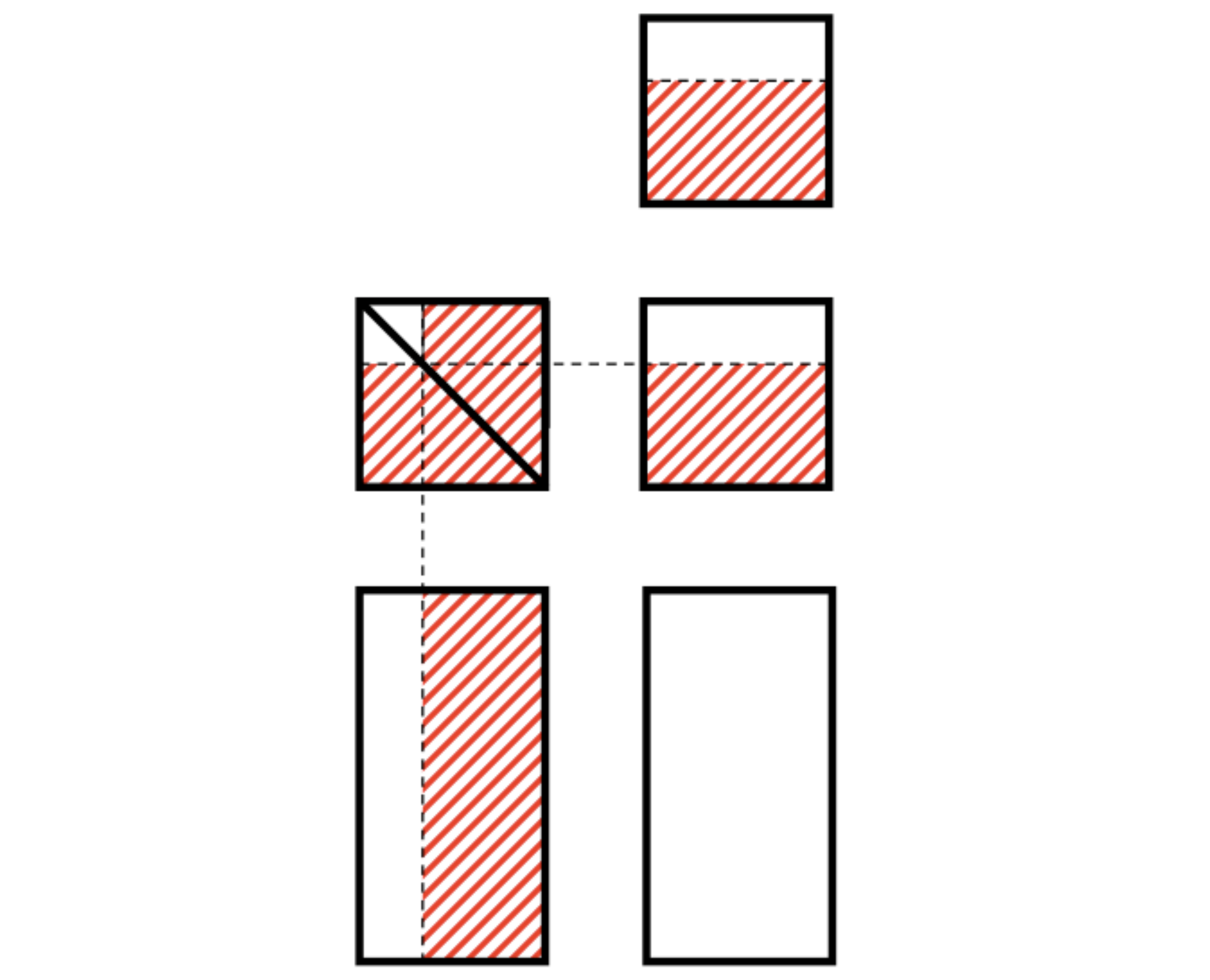}
\unitlength=1cm
\begin{picture}(5,0)
\put(1.00,0.15){$\bfUr$}
\put( 3.20,0.15){$\bfM$}
\put(1.00,4.75){$\bfWr$}
\put(2.10,6.10){$\bftVr$}
\put( 1.25,3.65){$\bf 0$}
\put( 1.73,4.05){$\bf 0$}
\put( 3.25,3.70){$\bf 0$}
\end{picture}
\caption{Graphical sketch of SVD truncation. The smallest singular values are replaced by zeros, thus the red hatched blocks of matrix $\bfW$ are made of zeros only. Therefore, the bottom part of the $\bfW\, \bftV$ matrix is also made of zeros. Being multiplied by zeros, the bottom part of matrix $\bftV$ and the right part of matrix $\bfU$ play no role. SVD truncation consists in removing the hatched areas from the matrices. The resulting reduced matrices are $\bfUr$, $\bfWr$ and $\bftVr$.}
\label{Fig graph_reduc}
\end{center}
\end{figure}
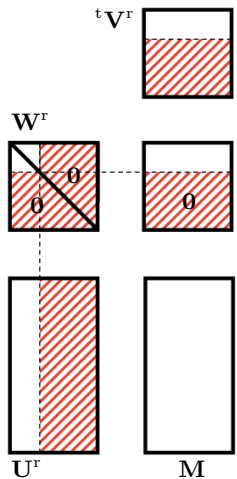

The handling of the matrices in SVD truncation is illustrated in fig. \ref{Fig graph_reduc} which shows graphically the matrix product of eq. (\ref{SVD}). The first step of the method consists in truncating the system, keeping only the relevant $r$ largest singular values. The red hatched part of the $\bfW$ matrix is now full of zeros. Hence, being multiplied by zeros, the lower part of the $\bftV$ and the right part of the $\bfU$ matrices can also be discarded (hatched blocks). The resulting reduced matrices are noted $\bfUr$ , $\bftVr$ and $\bfWr$. As can be seen, this operation does not modify the size of the $\bfM$ matrix. The goal here is to improve its conditioning. 

Doing this, we have, of course, introduced a systematical bias on the solution but, at the same time, we have reduced the uncertainty on the solution. The optimum number of discarded singular values can be defined as the value corresponding to the minimum average square error on the hit location. This value can only be found empirically, varying $r$ and calculating the solutions for a set of known locations. In fact, in most pulse shape analysis problems of germanium detectors, a large proportion of the singular values can be set to zero. Indeed, it has been observed that increasing the sample duration a lot has little effect on the precision of the hit location. The autocorrelation between successive samples is very strong ($\gamma_{1}/\gamma_{0}\approx  0.99$ where $\gamma_{k}$ is the $k^{\rm th}$ order autocorrelation coefficient), thus grouping the samples does not affect the information. The number of relevant samples being small, the number of relevant singular values is small as well.

Another positive effect of setting to zero singular values is that the number of lines in eq. (\ref{tVe=sr}) is lowered (see fig. \ref{Fig tVe}), which reduces drastically the computer time necessary to solve it. Indeed, as shown in fig. \ref{Fig graph_reduc}, the last lines of $\bftV$ can be discarded from the matrices. This corresponds to a projection of the signal samples onto an optimum subspace spanned by the column-vectors of $\bftVr$ that permits to keep a maximum information using a minimum number of components \cite{Des1}.

\subsection{Example of application}

\begin{figure}[htbp]
\begin{center}
\includegraphics[width=7.5cm]{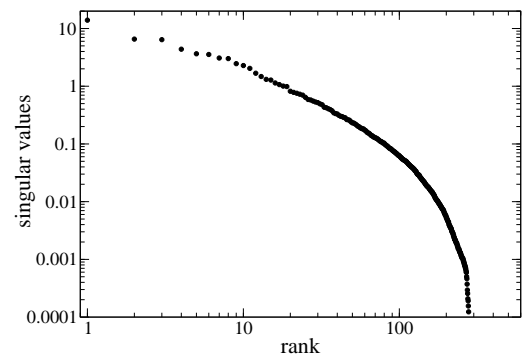}
\caption{Distribution of the singular values for a single AGATA detector segment for a 2~mm cubic grid.}
\label{Fig singval_M13}
\end{center}
\end{figure}

In order to illustrate the method, we have applied it to determine the position of single hits in a given segment of an AGATA crystal. The basis signals as well as the test signals are given by the MGS simulation. The test signals are altered by adding different amounts of noise and by a 3~ns standard deviation time shift jitter. A typical meta-signal is shown in fig. \ref{Fig meta-signal}. The signal basis ($\bfM$ matrix) is made of 950 meta-signals of 468 samples corresponding to a 2~mm cubic grid covering the volume of the segment. Its condition number being of the order of $10^{51}$, the direct solving of the system induces large uncertainties on the location of the hits. The distribution of the singular values shown in Fig \ref{Fig singval_M13} has no abrupt gap, thus the optimum number of discarded singular values has to be determined empirically.

The truncated system, eq. (\ref{tVe=sr}) and fig. \ref{Fig graph_reduc}, reads:
\begin{equation}
\label{tVre=sr}
\bftVr\,\bfe = \bfsr\quad {\rm with}\quad \bfsr={\bfWr}^{-1}\,\bftUr\,\bfs\,.
\end{equation}

\begin{figure}[htbp]
\begin{center}
\includegraphics[width=7.5cm]{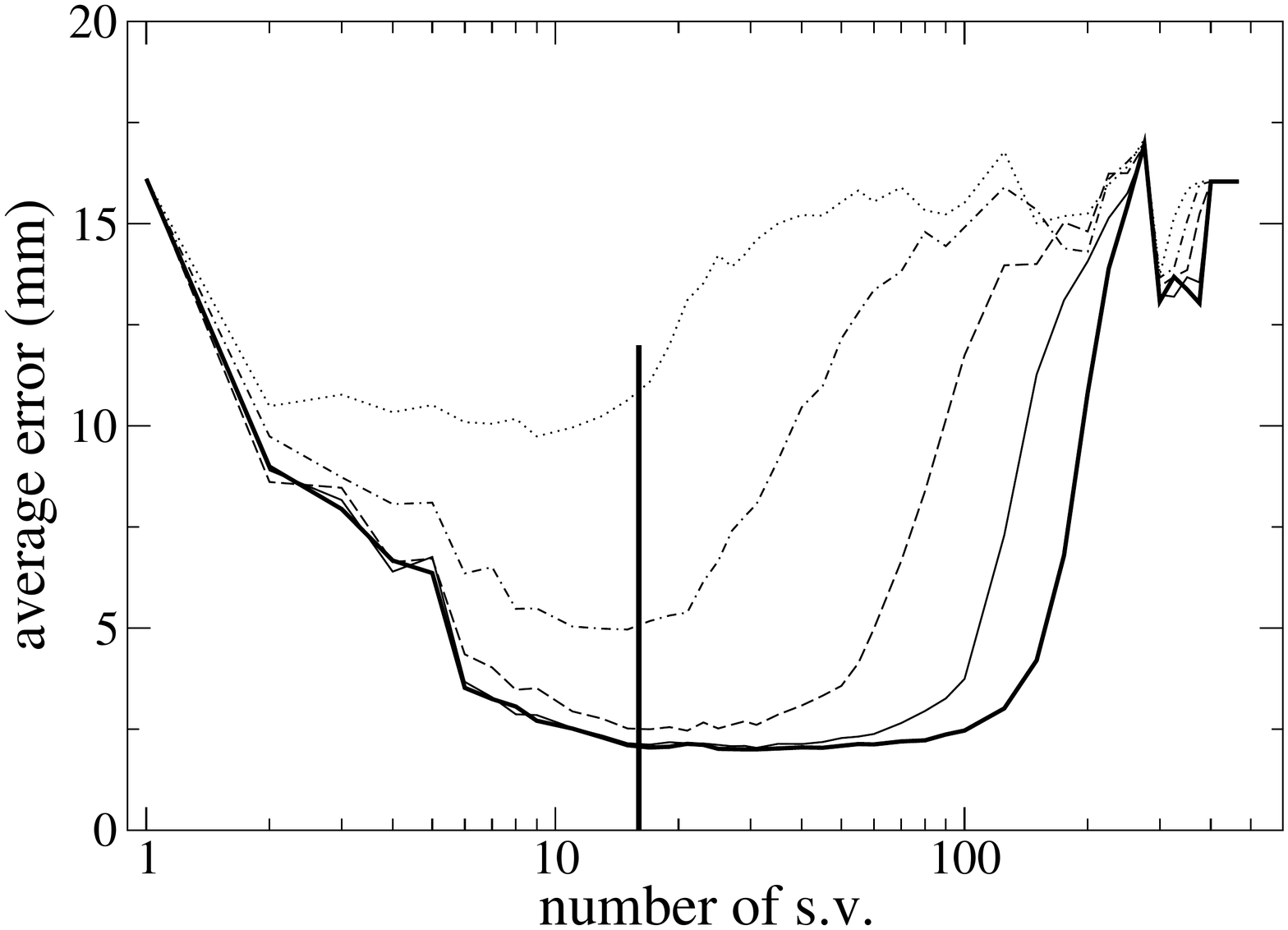}
\includegraphics[width=7.5cm]{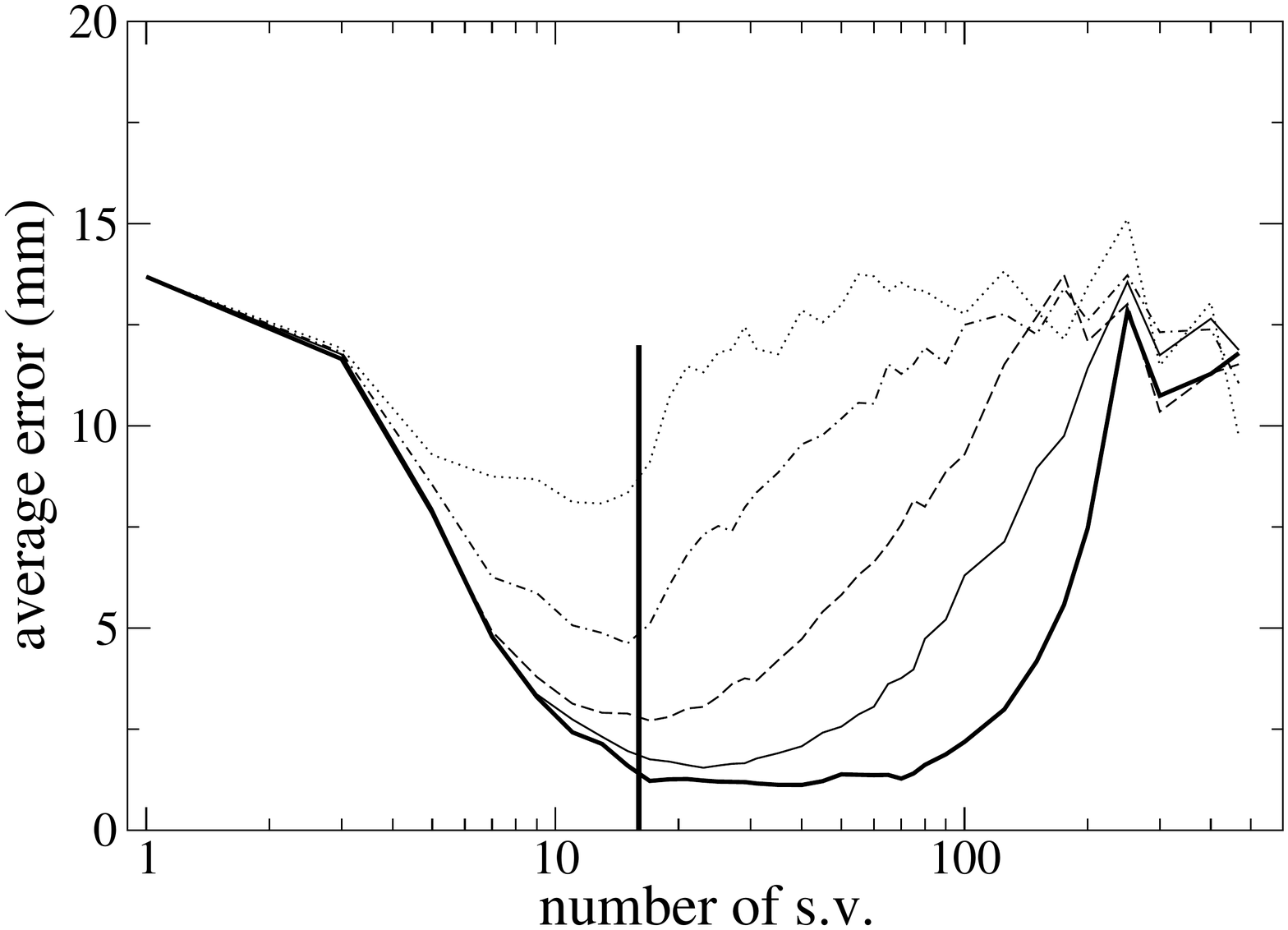}
\caption{Average error on the hit location as a function of the number of retained singular values, in the case of single hits at random positions in a given segment of an AGATA detector. Top panel: grid search; bottom panel: NNLS minimization. The line styles correspond to different amount of noise. Bold: $SNR=42.4$~dB, thin: $SNR=32.4$~dB, dashed: $SNR=22.4$~dB, dot-dashed: $SNR=12.4$~dB, dotted: $SNR=2.4$~dB. A 3~ns time jitter is also applied to the test signals. The vertical bar corresponds to 16 singular values.}
\label{Fig ed_sv_seg13}
\end{center}
\end{figure}

\noindent This equation can be solved in several ways. One is the use of a fast iterative inversion algorithm. We have used NNLS. Another method consists in comparing the reduced test signal to all the signals of the truncated basis of reduced signals $\bftVr$ (grid search on reduced signals). Here, the criterion for the best match is the residue \cite{Des5} (the scalar product criterion \cite{Dox} results in a faster, but slightly less precise, algorithm). The results obtained with both methods and for different signal-to-noise ratios are shown in fig.~\ref{Fig ed_sv_seg13}. 

In every case, the errors are large when only a few singular values are kept since the remaining information is too small to allow a precise localization, and when the number of singular values is large since the condition number is large. The optimum number of singular values depends slightly on the algorithm and on the signal-to-noise ratio. The resolution when 16 singular values are retained is close to the minimum, whatever the inversion method, the $SNR$, and for the time shift values that can be expected from electronics. It is remarkable that the optimum number of singular values is only twice the number of signals included in the meta-signals. For both algorithms, the computing time is proportional to the number of singular values. The SVD method, however, entails the calculation of the reduced meta-signal $\bfsr$ event by event. Nonetheless, keeping 16 singular values out of 468 reduces the computation time by a factor of about 20. 

This analysis also shows that grid search is well adapted to the determination of single hits, since it is the fastest algorithm\footnote{Only when the noise is low ($SNR > 18$~dB), the Grid Search solving of eq. (\ref{Me=s}) may give a slightly better resolution than eq. (\ref{tVre=sr}) truncated to 16 singular values, but the computing time is also 20 times longer.}. The inversion method developed in the next section gives better results when the signals, resulting from several hits, overlap.

\section{Signal decomposition algorithm}

\subsection{More than one hit in a single segment}
 
Depending on the type of application, the situations when a gamma interacts more than once in the same segment (or two gamma interact simultaneously in the same segment), are treated in different ways. For source location, or more generally, when the precision on the measurement is more important than the amount of analyzed photons, these events are simply discarded. However, even in this case, in order to be rejected, multi-hits have to be discriminated from single hits. 

In other cases, typically in nuclear physics experiments yielding high gamma-ray multiplicities, discarding events would introduce a crippling bias. Hence, whenever possible, every hit has to be located. Grid search algorithms are not well adapted to the solving of multi-hits. Indeed, for single hits the computing time is proportional to the number of grid points, for double hits this time is proportional to the square of the number of grid points times the number of possible energy sharing between the two hits. Thus the computing time increases more rapidly than the exponent of the number of hits. This drawback is also true for most artificial intelligence techniques. Using the matrix formalism, as will be seen now, a faster algorithm can be developed.

\subsection{Locations of the multi-hits}

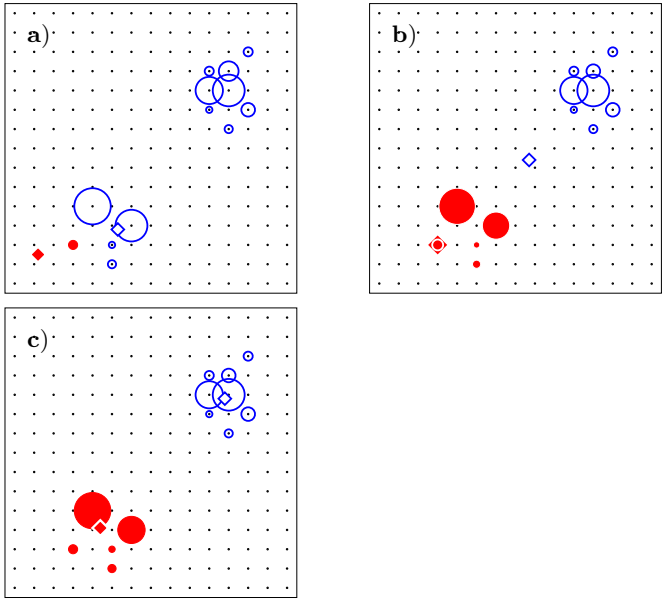
\begin{figure}[htbp]
\includegraphics[width=4.cm]{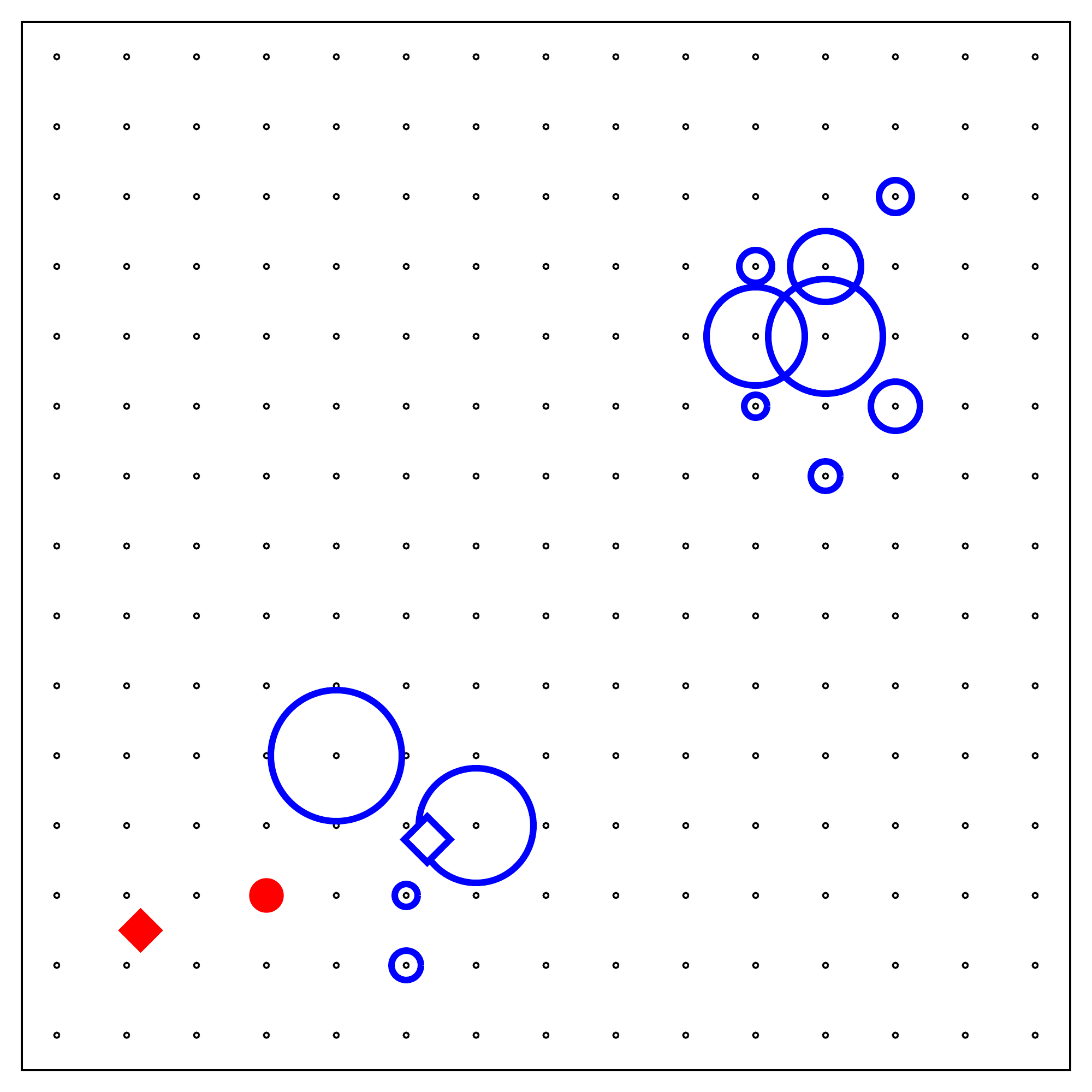}
\includegraphics[width=4.cm]{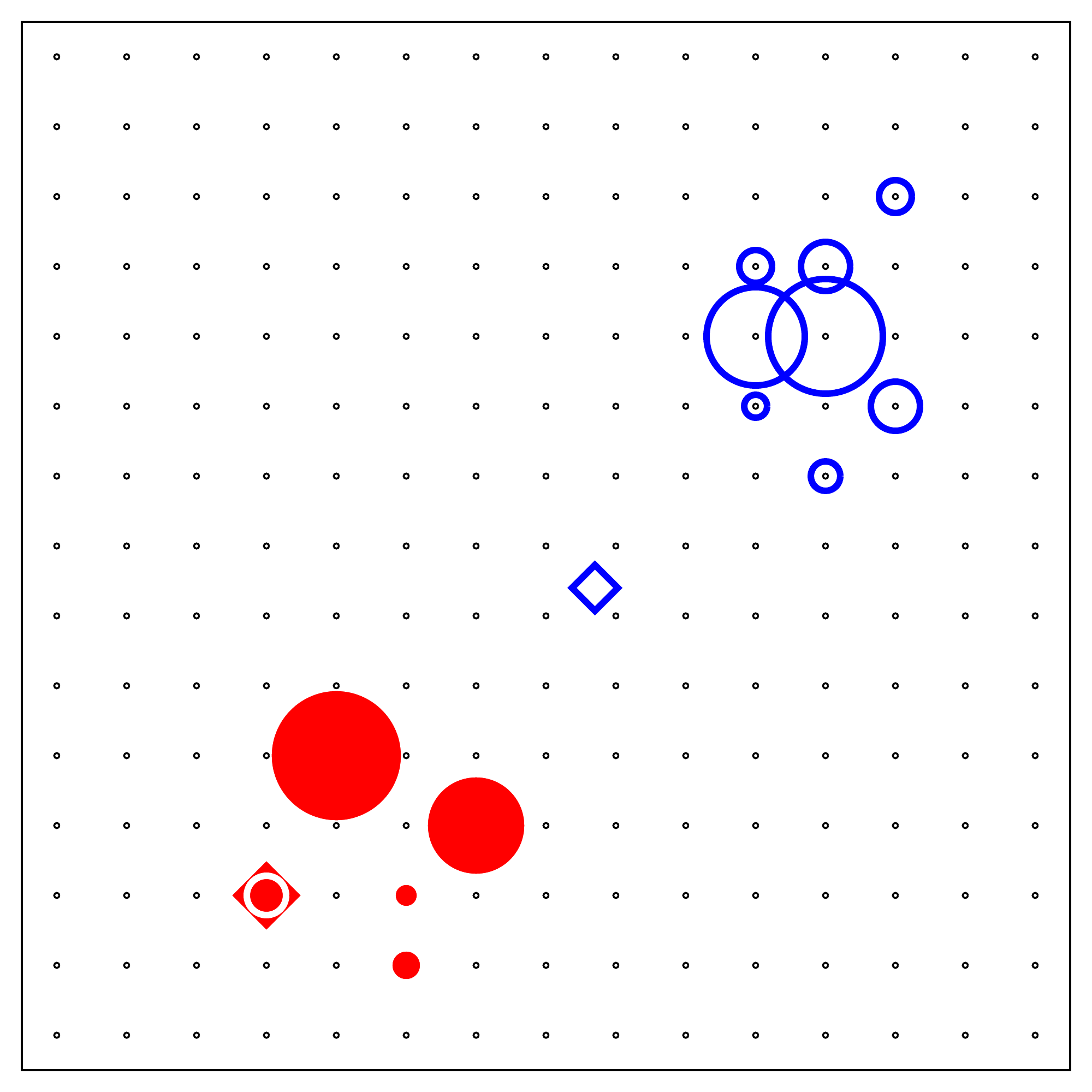}
\includegraphics[width=4.cm]{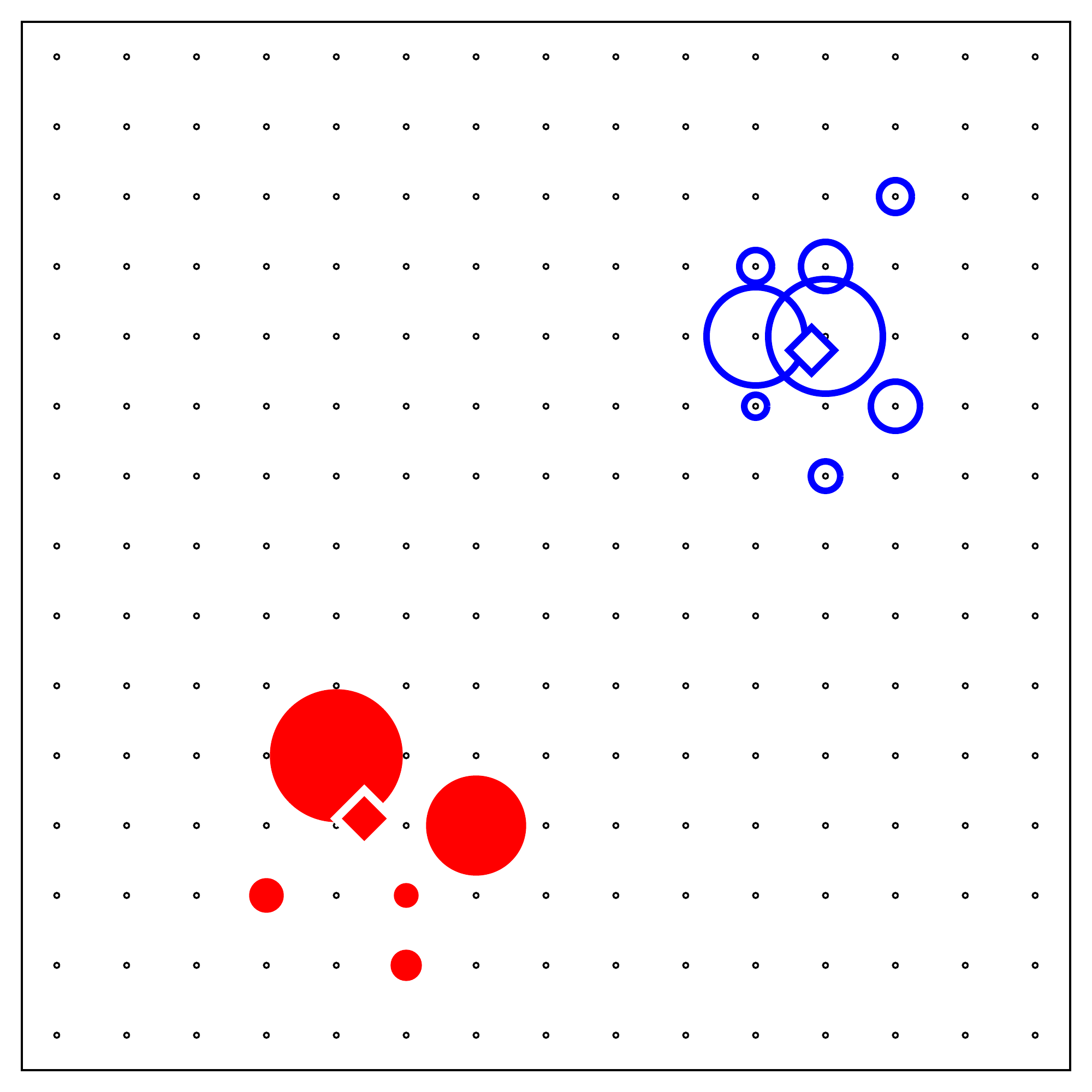}
\unitlength=1cm
\begin{picture}(5,0)
\put(-3.70,7.45){$\bf a)$}
\put( 1.10,7.45){$\bf b)$}
\put(-3.70,3.40){$\bf c)$}
\end{picture}
\caption{Illustration of the mobile center algorithm. The surfaces of the dots are proportional to the corresponding component in the $\bfe$ vector; their colors correspond to the closest mobile center. The two mobile centers are represented with diamonds. a) Initial (random) positions of the mobile centers, b) positions after one iteration, c) positions after two iterations (final positions) and estimated hit locations. }
\label{Fig Gi}
\end{figure}

Signal decomposition consists in solving eq. (\ref{tVre=sr}) in order to determine the location and the deposited energy for each gamma interaction. In the multi-hit case, each hit appears in the solution vector as a cloud of adjacent (in the position space) non-zero components (fig. \ref{Fig Gi}). If the clouds have strong overlaps then it is not possible to distinguish the different hits. This happens when the hits are too close or when they occur in a part of 
the segment where the sensitivity is low or when the energy deposits are low (see low $SNR$ part of fig.~\ref{Fig ed_rI_snr_t0}). When the clouds are separated, each hit location is estimated using eq. (\ref{G}) in which the sum runs only on the components of the corresponding cloud. Both the cloud discrimination and the hit location estimation can be computed rapidly using the mobile center algorithm. This method is illustrated in fig. \ref{Fig Gi}. The estimated locations of the hits are first sorted at random. Then each positive component of the solution matrix is associated to the closest of these centers and each center is replaced by the barycenter of its associated components. This procedure is repeated iteratively until the center locations are stable. Their final positions are the estimated locations of the hits. 

Most of the time, the main difficulty is to determine the number of simultaneous hits in a segment. Many a priori algorithms (in the sense that the number of hits is determined first) have been tested but the performances are still not very satisfactory. An interesting strategy is proposed in reference \cite{Cre1}. The matrix formalism allows an a posteriori determination. First, the linear system is solved. The total cloud of non-zero components can then be analyzed. Several strategies are possible for searching for independent sub-clouds corresponding to the different hits. We have found that the analysis of the cloud inertia tensor (multi-hits correspond to larger moments) was the most robust method in the case of coaxial detectors. 

An important advantage of the singular value decomposition method is that it reduces the dispersion of the clouds. This property facilitates the separation of the clouds, thus the identification of the hits. Yet, whatever the chosen method, it seems difficult to discriminate more than two hits in a segment and the errors on the location and the energy sharing are small only if the locations are not too close and the energies are not too low. 

\subsection{Protocol for the whole crystal}

We now consider the whole array of segments. Instead of being built with the signals of the hit segment and of its neighbors, the meta-signals are obtained by the concatenation of all the segment signals and the vector $\bfe$ gather the energy deposits on all the crystal grid voxels. This corresponds to a large increase of the transformation matrix size. However, using SVD truncation, the resulting size is not excessive for on-line applications. 

\begin{figure}[htbp]
\begin{center}
\includegraphics[width=7.5cm]{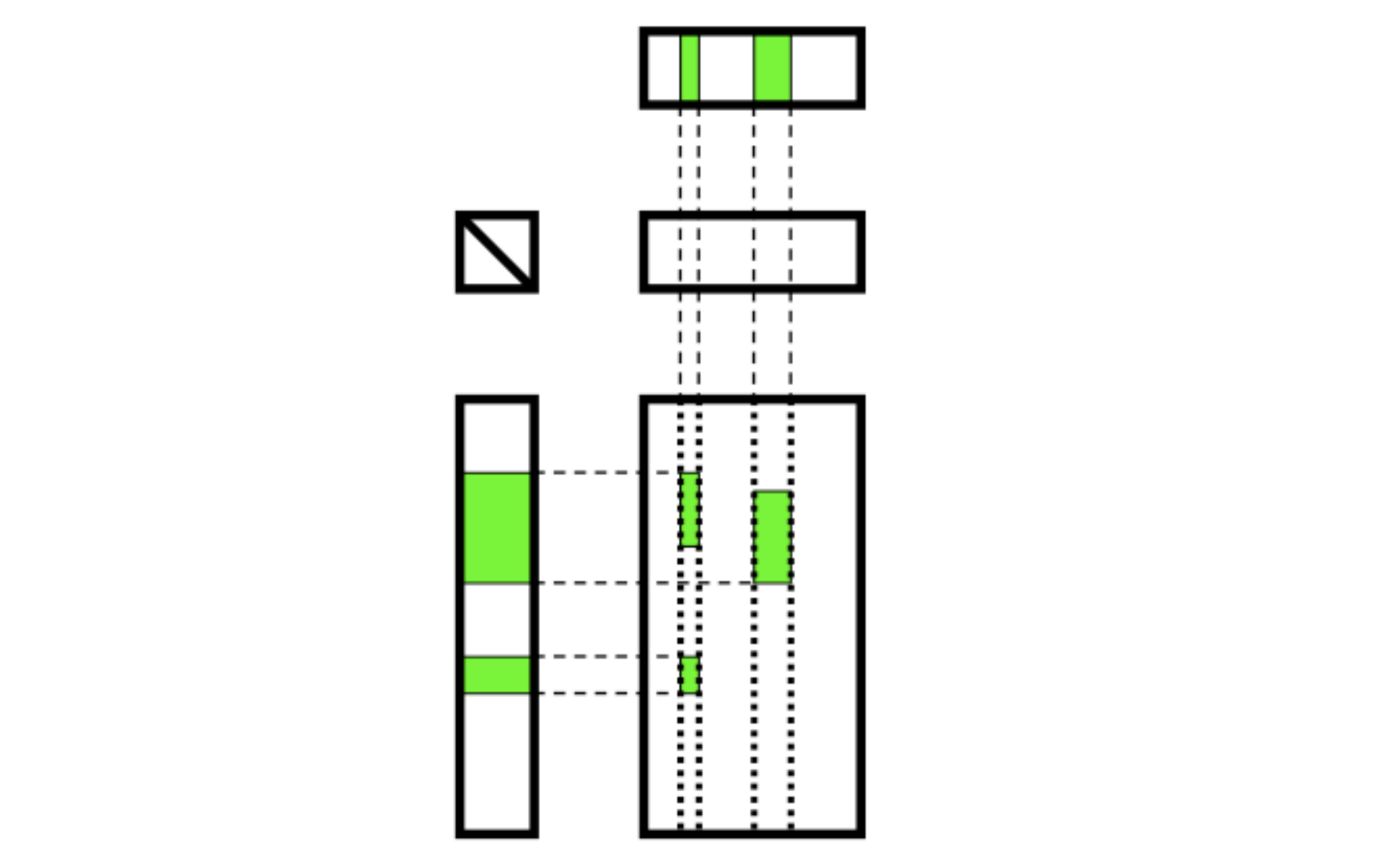}
\unitlength=1cm
\begin{picture}(5,0)
\put( 1.20,0.13){$\bfUr$}
\put( 2.55,0.13){$\bfM$}
\put( 0.60,3.45){$\bfWr$}
\put( 2.50,4.90){$\bftVr$}
\end{picture}
\caption{Graphical sketch of singular value decomposition after truncation (the red hatched blocks of fig. \ref{Fig graph_reduc} are discarded). In typical events, not all the segments are hit simultaneously, thus, only the columns of $\bfM$ (dotted line) and $\bftVr$ corresponding to the hit segments have to be kept for the location of the hits. Moreover, when given segments are hit, signals are seen only for these segments and their neighbors. Thus, only the corresponding lines of the $\bfM$ and $\bfUr$ matrices have to be used in the signal decomposition (green blocks). }
\label{Fig hit}
\end{center}
\end{figure}
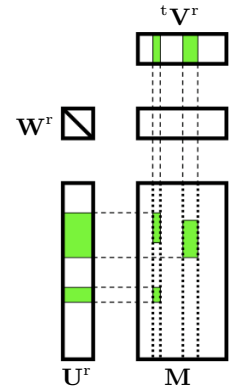

In a typical gamma event, only a small number of segments are hit, thus it is not necessary to solve the linear system using the whole $\bfM$ and $\bfe$ matrices. Only the coordinates of the unknown vector corresponding to grid points inside the hits segments, and the corresponding columns of the basis signal matrix, have to be retained. In fig. \ref{Fig hit}, we consider a case where the hit segments correspond to the two columns indicated by dotted lines in matrix $\bfM$. The interactions induce signals only in the hit segment and its neighbors. The useful part of the meta-signals are indicated by the green blocks in the $\bfM$ matrix. Thus, we introduce the $\bftVhit$ and $\bfUhit$ matrices which are built only with the useful blocks. Similarly, the $\bfehit$ vector is composed only with the grid points belonging to the hit segments (green columns of  $\bftVr$) and $\bfshit$ is composed by the signals of the hit segments and their neighbors (green lines of $\bfUr$).

\begin{figure}[htbp]
\begin{center}
\includegraphics[width=7.5cm]{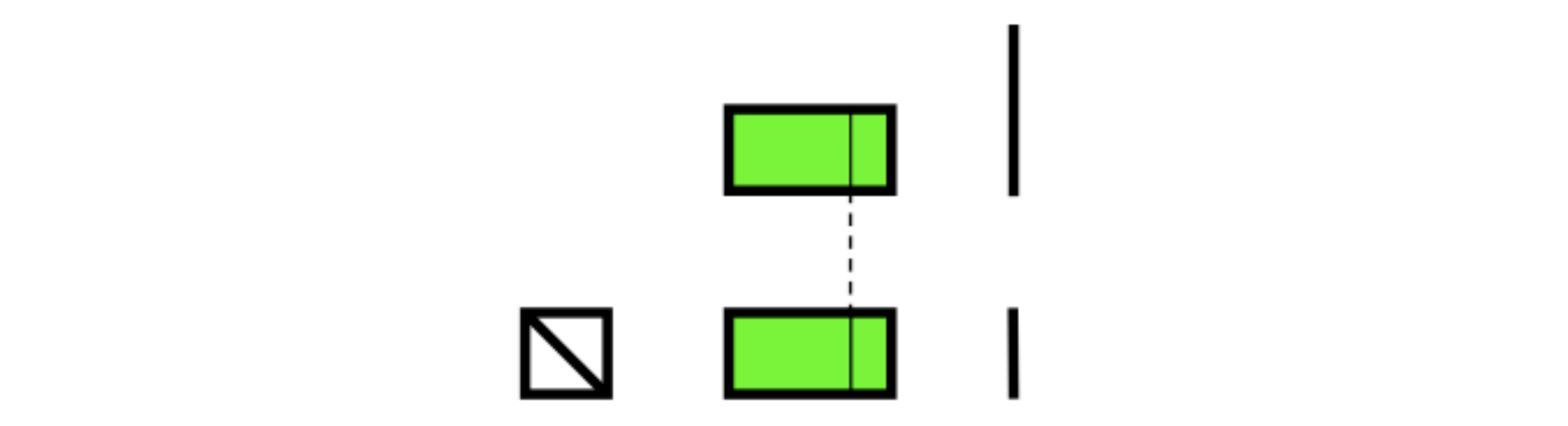}
\unitlength=1cm
\begin{picture}(5,0)
\put( 3.85,1.85){$\bfshit$}
\put( 3.85,0.70){$\bfsrhit$}
\put( 1.17,0.07){$\bfWr$}
\put( 1.67,0.12){$^{-1}$}
\put( 2.30,2.15){$\bftUhit$}
\put( 2.40,0.07){$\bfRhit$}
\end{picture}
\caption{Graphical sketch of the construction of the $\bfsrhit$ reduced signal corresponding to the right-hand side of eq. (\ref{tVre=sr}).}
\label{Fig srhit}
\end{center}
\end{figure}
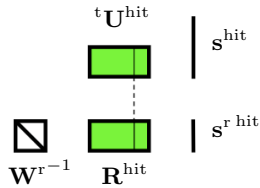

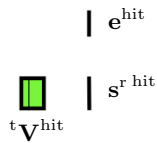
\begin{figure}[htbp]
\begin{center}
\includegraphics[width=7.5cm]{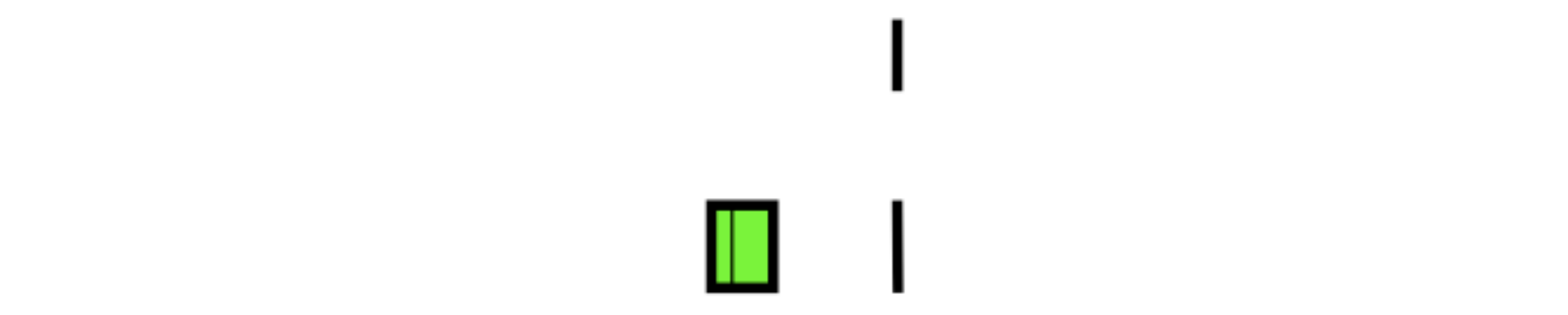}
\unitlength=1cm
\begin{picture}(5,0)
\put( 3.30,1.55){$\bfehit$}
\put( 2.00,0.07){$\bftVhit$}
\put( 3.30,0.65){$\bfsrhit$}
\end{picture}
\caption{Graphical sketch of eq. (\ref{tVre=sr}). The $\bftVhit$ matrix is the concatenation of the useful (green) blocks of matrix $\bftVr$. Only the components of vectors $\bfe$ corresponding to the grid points which are inside one of the hit segments, are retained.}
\label{Fig tVhitehit}
\end{center}
\end{figure}

As we have seen, the system to be solved is given by eq.  (\ref{tVre=sr}) and fig. \ref{Fig tVe}. The $\bfRhit$ matrix is calculated first (fig. \ref{Fig srhit}). The final linear system is shown in fig. \ref{Fig tVhitehit}. As can be seen, the size of the system has been drastically reduced. 

In order to use this method in on-line applications, it is important to pre-calculate as many matrices as possible. The singular value decomposition can be done off-line and the resulting $\bftVr$ and $\bfR$ matrices are memorized. Event by event, the $\bftVhit$ and $\bfRhit$ matrices have to be extracted from the previous matrices and the reduced signal is obtained by multiplying the detected signal by the $\bfRhit$ matrix. Finally, only the small $\bftVhit\ \bfehit = \bfsrhit$ system is to be solved. 

One advantage of this protocol is that the singular value decomposition has to be done only once, off-line. Thus, on an event-by-event basis, one has only to select the useful parts of the matrices and solve a small system inversion. 

A more rigorous, but more complex, procedure is first to select the useful blocks then to the calculate the singular value decomposition. This protocol is developed in Appendix B.

\section{Conclusion}

The response function of High Purity Ge detectors has several characteristics that make signal decomposition difficult. The relation between the pulse shapes and the locations of the hits is not always bijective, the response function amplifies the signal noise to the hit location estimation, signal decomposition is an ill-posed problem and the size of the linear system to be solved is very large.

These problems have been addressed using the matrix formalism. The first advantage of this method is to allow a mathematical analysis of the response function of the individual segments or of the whole germanium
crystal. Using the SVD analysis, we have evaluated how signal uncertainties alter the precision on the estimated hit location. This decomposition also indicates the maximum number of grid points that a segment or the crystal can accommodate. We then discuss the SVD truncation leading to the reduction of the size of the system, which permits both to reduce the computing time and to decrease the uncertainty on the hit location estimation. 

For high-rate on-line applications, the SVD of the whole transformation matrix is computed off-line. Each event has to be solved selecting only the relevant blocks of the matrices. 

In conclusion, whatever the algorithm used for signal decomposition, the signal basis should be analyzed and reduced using SVD in order to speed up the on-line computations and to improve the conditioning of the response function. For complex events, that is when the detected signal results from multiple gamma interactions, the matrix method gives very good results in terms of energy and position precision as well as of computing time. 

\appendix

\section{Appendix A: Box diagram for matrix products}

For a better readability, the matrix handling of this paper are illustrated by box diagrams. This Appendix shows how such diagrams are constructed. A matrix is represented by a box which height is proportional to the number of lines and which width is proportional to the number of columns. The product of two matrices is represented in the following way:
\begin{figure}[hhtbp]
\begin{center}
\includegraphics[width=5cm]{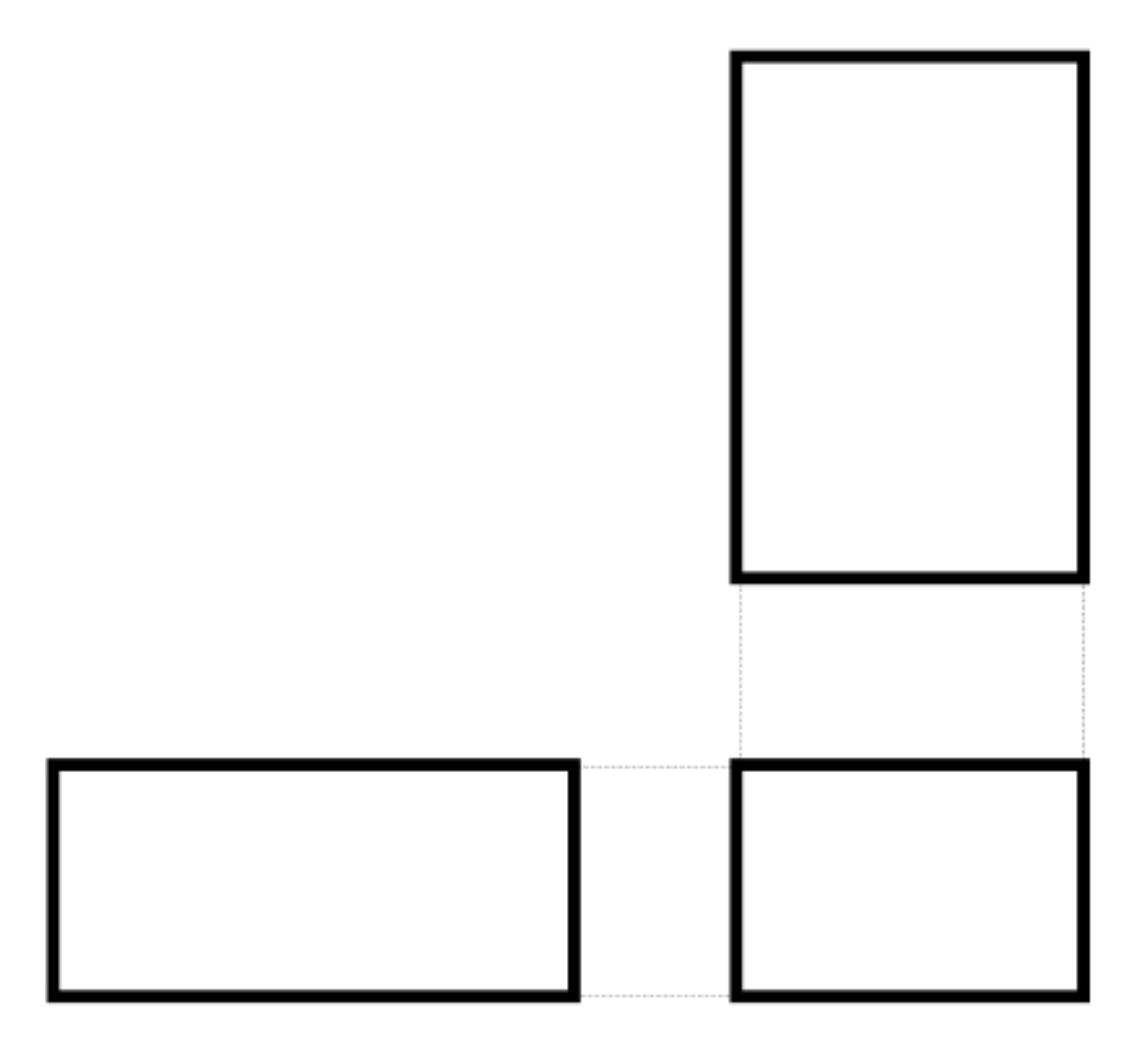}
\unitlength=1cm
\begin{picture}(5,0)
\put( 1.25,0.20){$\bfA$}
\put( 2.75,3.60){$\bfB$}
\put( 3.80,0.20){$\bfC$}
\put( 0.33,1.00){$a_{i1}$}
\put( 2.00,1.00){$a_{in}$}
\put( 1.15,1.00){\ldots}
\put( 4.00,3.50){\vdots}
\put( 3.90,4.40){$b_{1j}$}
\put( 3.90,2.60){$b_{nj}$}
\put( 3.90,1.00){$c_{ij}$}
\thinlines
\put( 2.50,1.05){\vector(1,0){1.3}}
\put( 4.00,2.47){\vector(0,-1){1.22}}
\end{picture}
\caption{Graphical sketch representing the product $\bfA\ \bfB=\bfC$.}
\label{Fig ABC}
\end{center}
\end{figure}

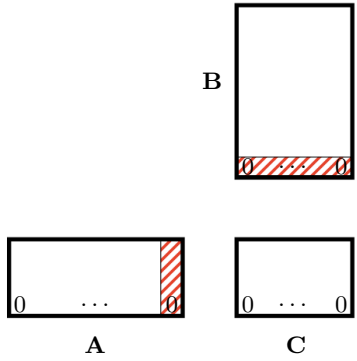
\begin{figure}[htbp]
\begin{center}
\includegraphics[width=5cm]{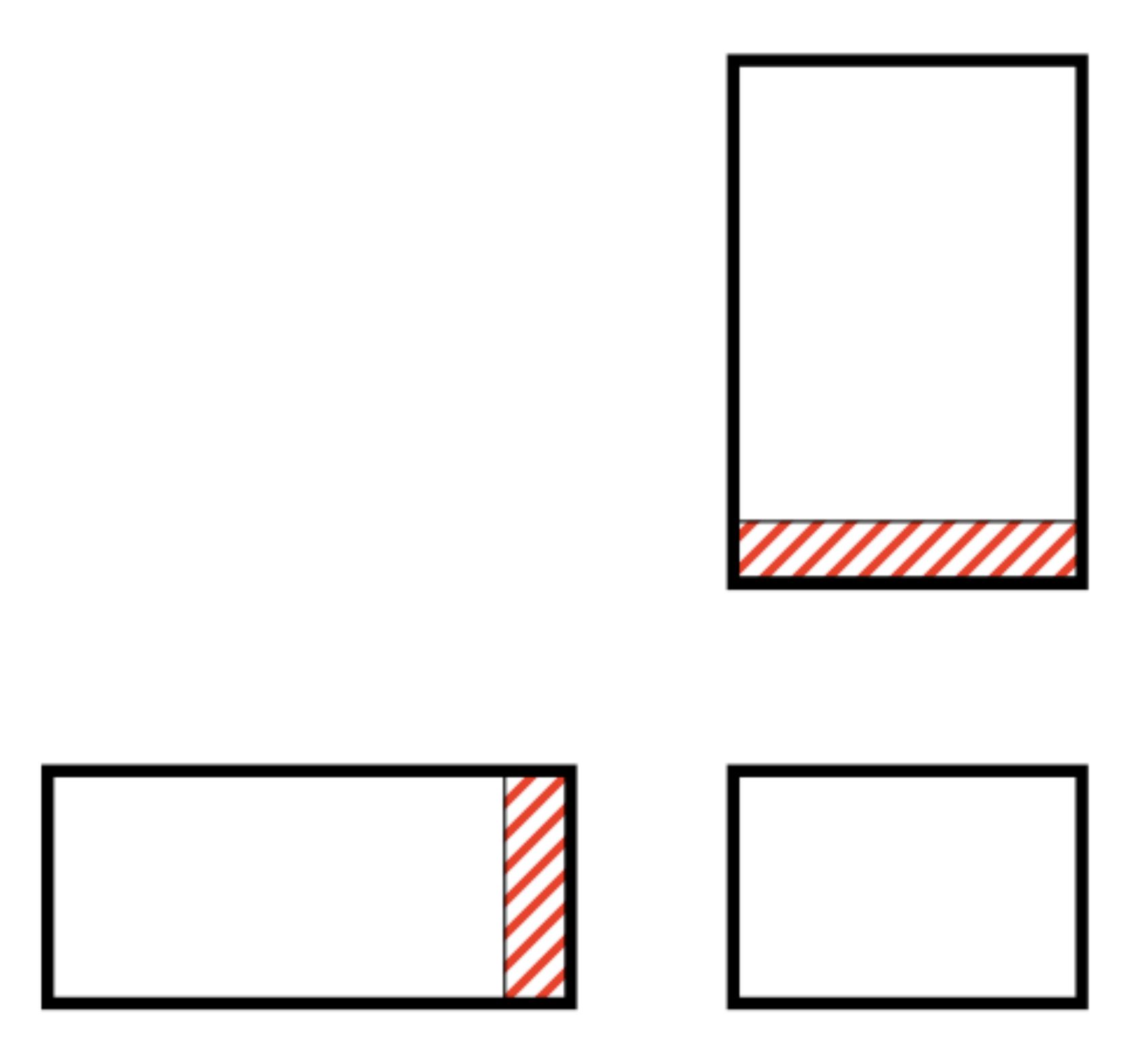}
\unitlength=1cm
\begin{picture}(5,0)
\put( 1.20,0.07){$\bfA$}
\put( 2.75,3.57){$\bfB$}
\put( 3.85,0.07){$\bfC$}
\put( 0.27,0.62){0}
\put( 2.27,0.62){0}
\put( 1.15,0.69){\ldots}
\put( 3.75,2.52){\ldots}
\put( 3.75,0.69){\ldots}
\put( 3.27,2.45){0}
\put( 4.50,2.45){0}
\put( 3.27,0.62){0}
\put( 4.50,0.62){0}
\end{picture}
\caption{Illustration of matrix products properties using a box diagram. The red hatched blocks of $\bfA$ and $\bfB$ can be discarded.}
\label{Fig discard}
\end{center}
\end{figure}

The $c_{ij}$ component is the sum-product of the line of $\bfA$ and the column of $\bfB$ pointing towards it. The vectors, assimilated to single column matrices, are simply represented by a vertical line as in fig. \ref{Fig Me}. Some properties of matrix products can be easily visualized using box diagrams. For example, as shown in fig. \ref{Fig discard}, if the last line of $\bfA$ is made of zeros, then the last line of $\bfC$ is also made of zeros. If the last line of $\bfB$ is made of zeros then the removals of the last column of $\bfA$ and of the last line of $\bfB$ do not modify $\bfC$.

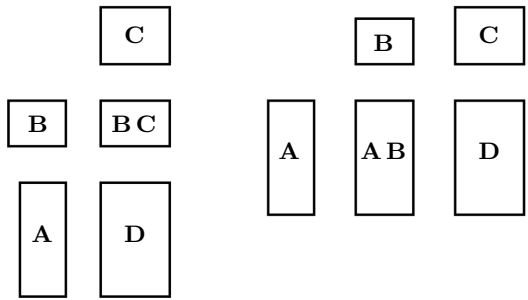
\begin{figure}[htbp]
\begin{center}
\includegraphics[width=7cm]{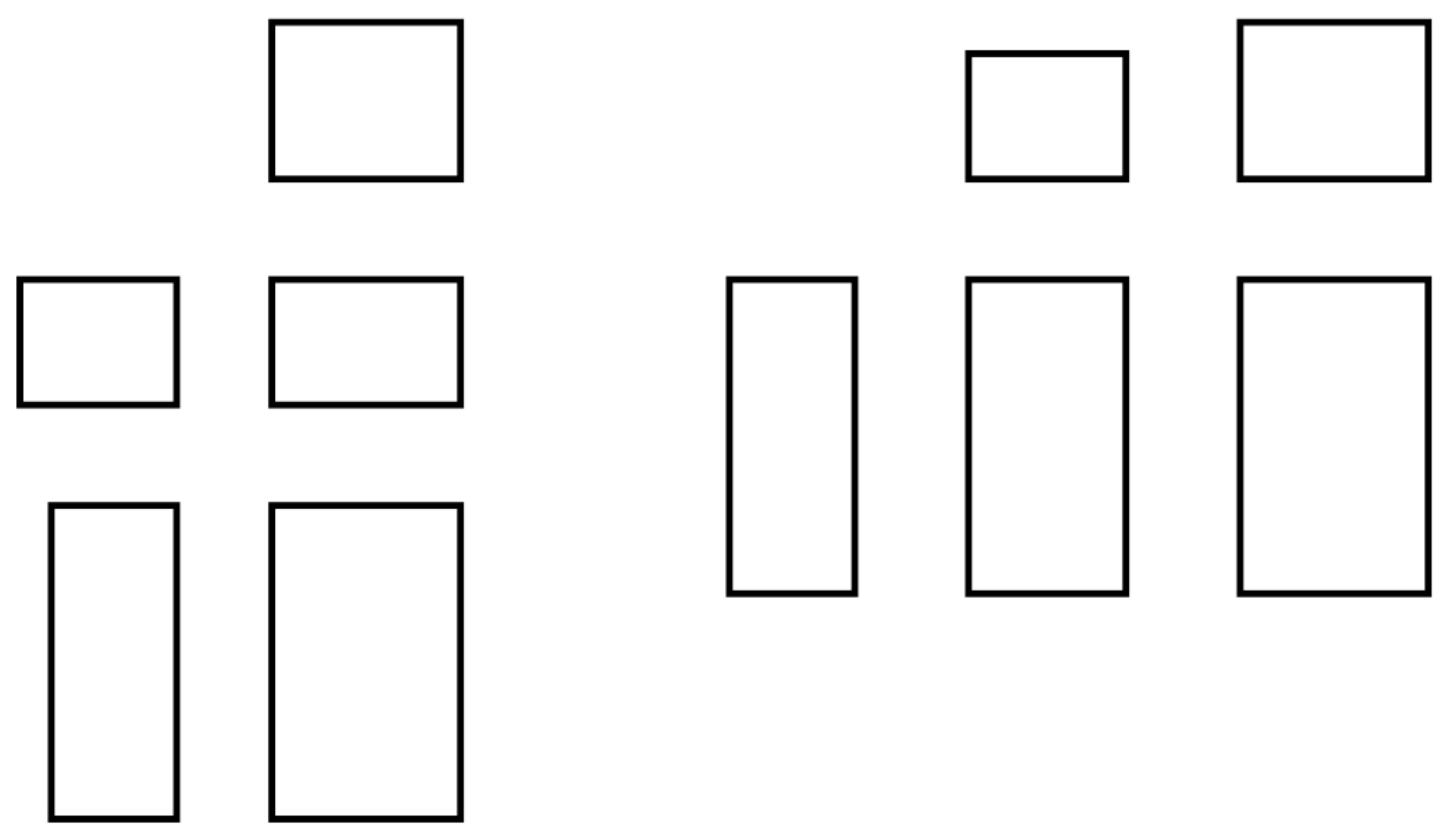}
\unitlength=1cm
\begin{picture}(5,0)
\put(-0.60,1.20){$\bfA$}
\put( 0.62,1.20){$\bfD$}
\put(-0.65,2.65){$\bfB$}
\put( 0.45,2.65){$\bfB\,\bfC$}
\put( 0.62,3.83){$\bfC$}

\put( 2.65,2.30){$\bfA$}
\put( 3.73,2.30){$\bfA\,\bfB$}
\put( 5.28,2.30){$\bfD$}
\put( 3.90,3.73){$\bfB$}
\put( 5.28,3.83){$\bfC$}
\end{picture}
\caption{Diagrams for the product of three matrices: $\bfA\ \bfB\ \bfC=\bfD$.}
\label{Fig ABCD}
\end{center}
\end{figure}

The product of three matrices can be represented by two different diagrams (fig. \ref{Fig ABCD}).

\section{Appendix B: Protocol for subsets of segments}

\subsection{Splitting of the linear system}

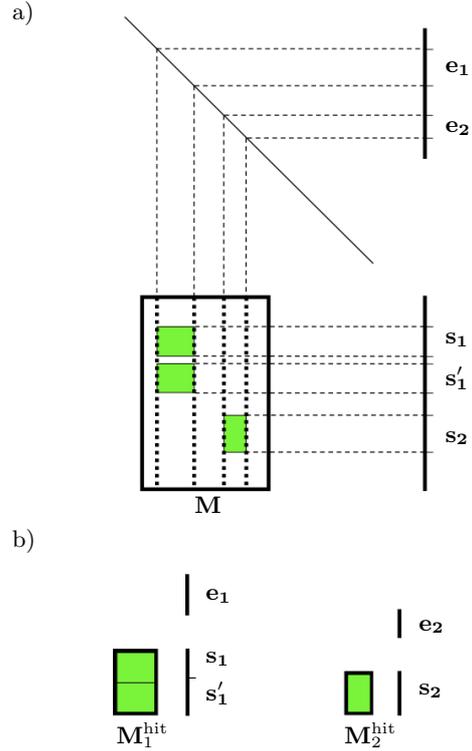
\begin{figure}[htbp]
\begin{center}
\includegraphics[width=7.5cm]{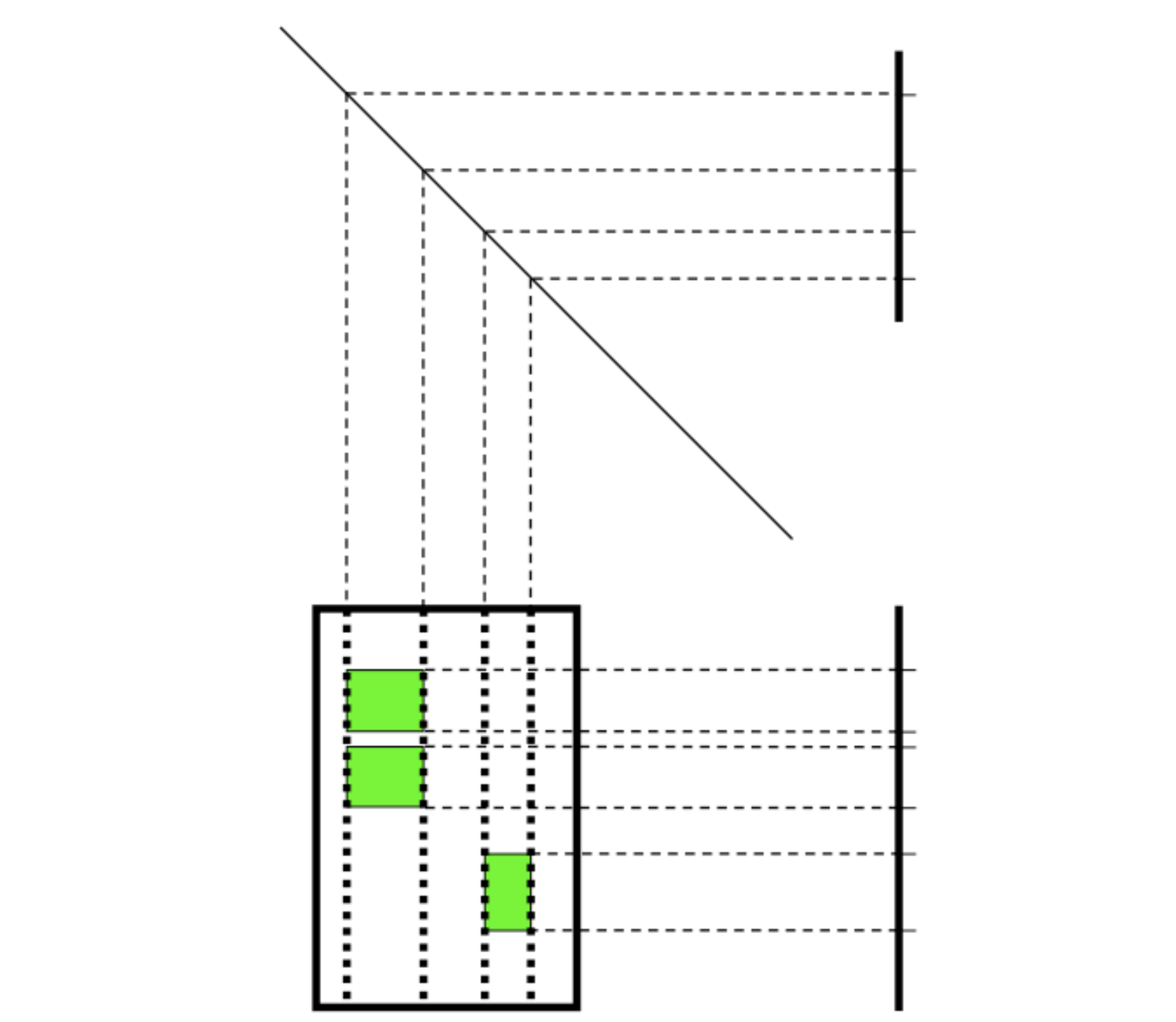}
\vspace{.8 cm}

\includegraphics[width=7.5cm]{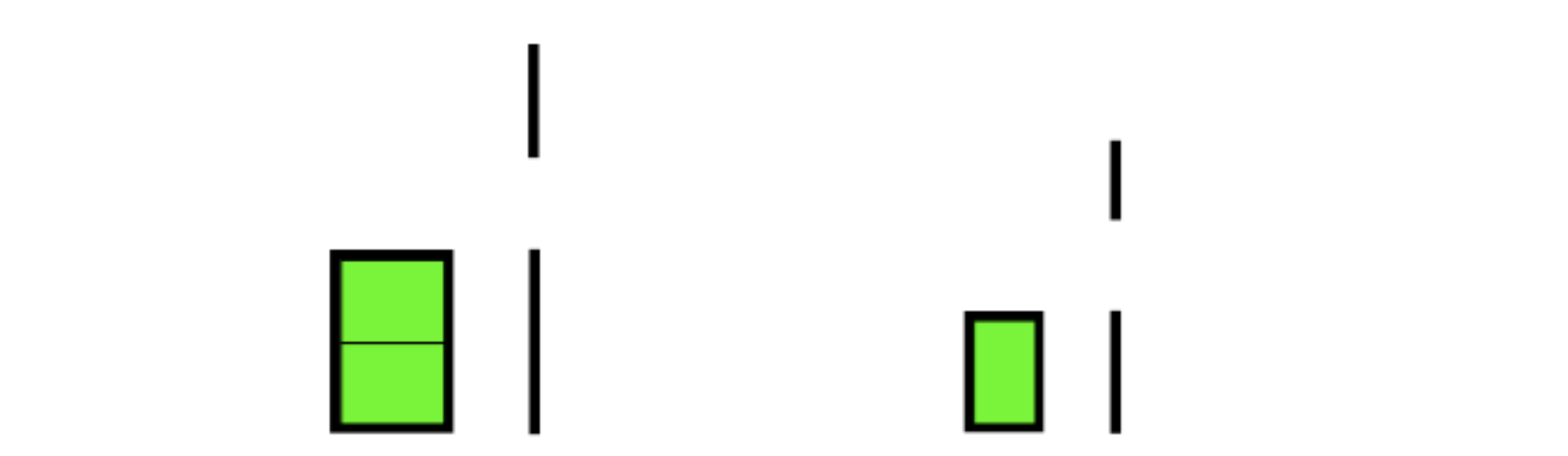}
\vspace{.3 cm}
\unitlength=1cm
\begin{picture}(5,0)
\put(-1.00, 9.70){a)}
\put( 4.70, 9.00){$\bf e_{1}$}
\put( 4.70, 8.20){$\bf e_{2}$}
\put( 4.70, 5.40){$\bf s_{1}$}
\put( 4.70, 4.85){$\bf s_{1}'$}
\put( 4.70, 4.10){$\bf s_{2}$}
\put( 1.40, 3.15){$\bfM$}

\put(-1.00, 2.70){b)}
\put( 1.55, 2.00){$\bf e_{1}$}
\put( 4.35, 1.60){$\bf e_{2}$}
\put( 1.55, 1.17){$\bf s_{1}$}
\put( 1.55, 0.67){$\bf s_{1}'$}
\put( 4.35, 0.75){$\bf s_{2}$}
\put( 1.30, 0.96){\_}
\put( 0.37, 0.10){$\bfMhit_{1}$}
\put( 3.37, 0.10){$\bfMhit_{2}$}
\end{picture}
\caption{Splitting of the linear system into two independent smaller systems. a) Selection of the useful blocks of the initial $\bfM$ matrix (no block overlap as in fig. \ref{Fig hit}). b) Resulting two independent linear systems.}
\label{Fig split}
\end{center}
\end{figure}

An event is often composed of groups of hits which happen in well separated zones of the detector, in that sense that the signals induced by each group do not overlap with the signals induced by the others. In that case, every group of hits can be treated independently. fig. \ref{Fig split} shows how the matrix system can be split into two separated systems.

\subsection{Inverse protocol}

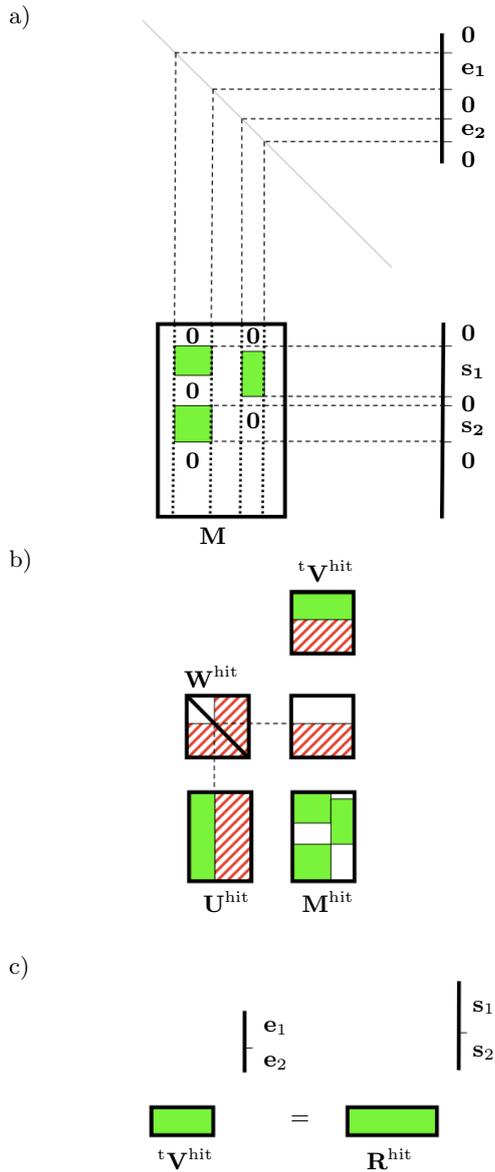
\begin{figure}[htbp]
\begin{center}
\includegraphics[width=7.5cm]{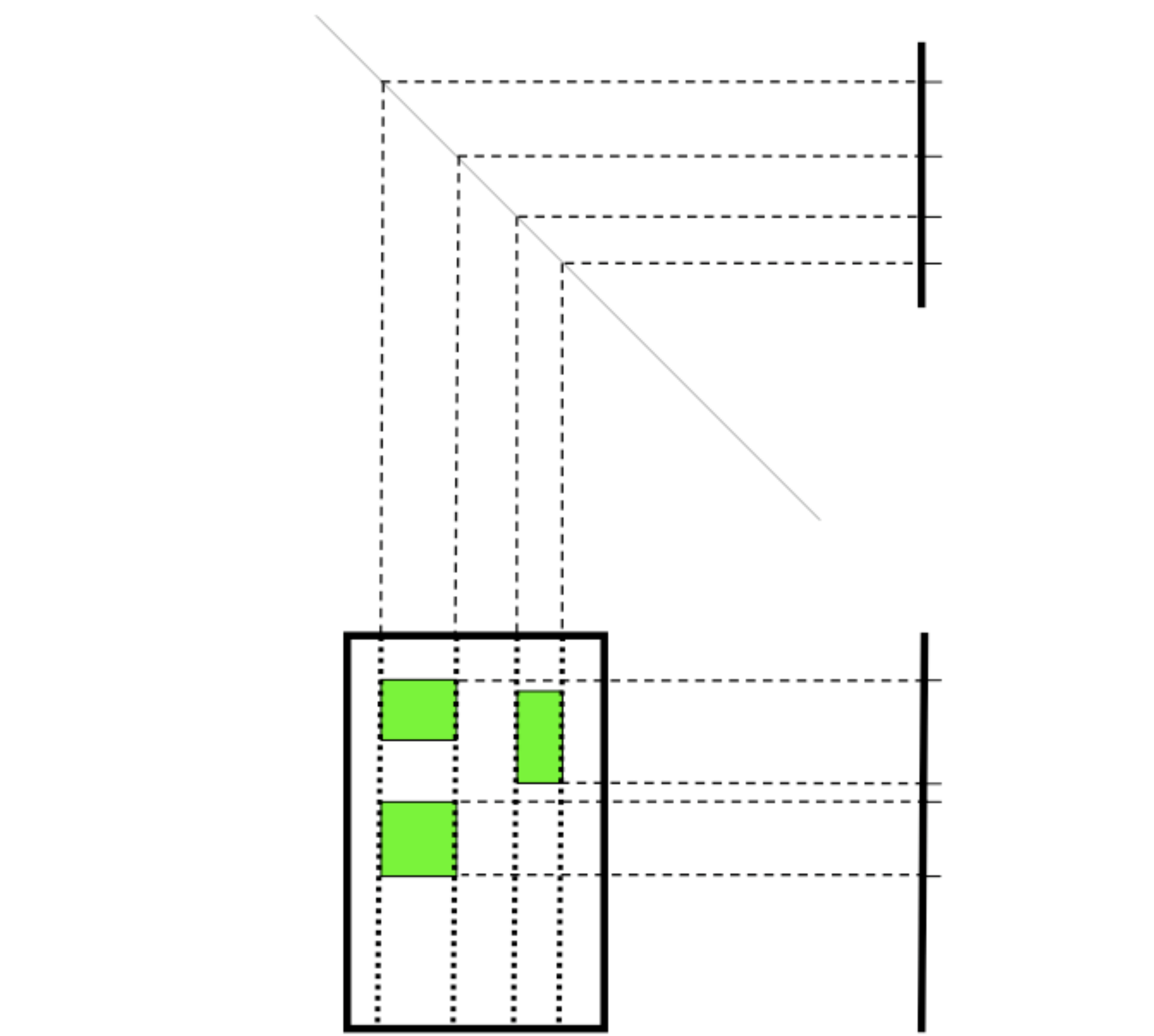}
\vspace{.6 cm}

\includegraphics[width=7.5cm]{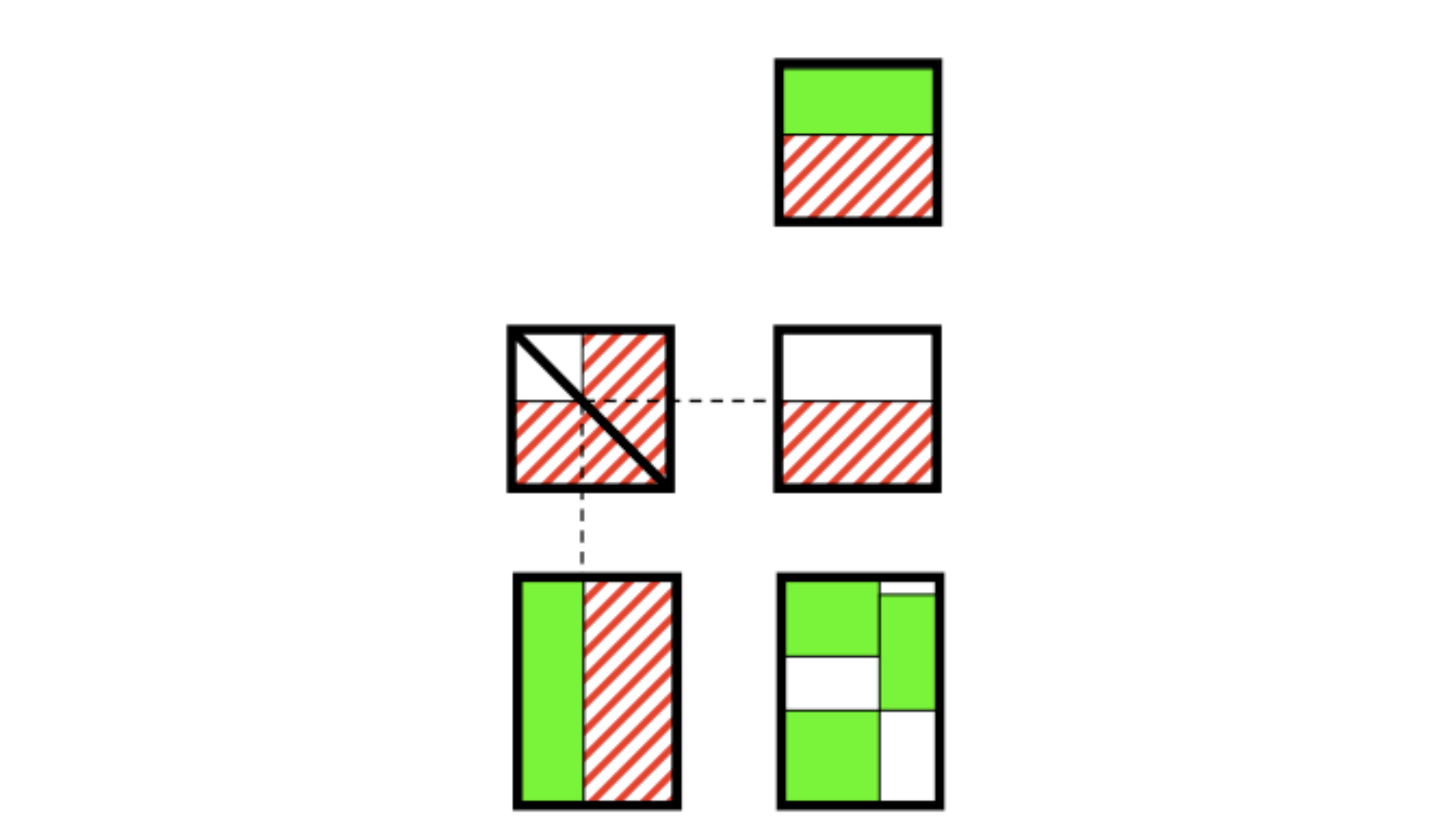}
\vspace{1. cm}

\includegraphics[width=7.5cm]{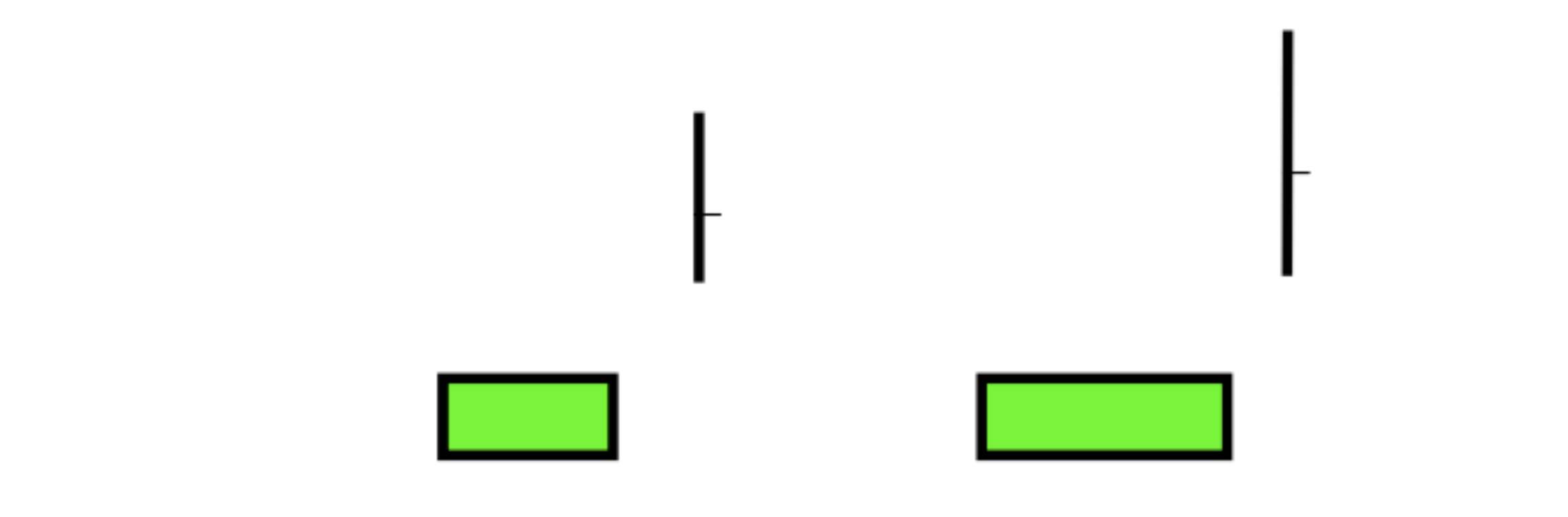}
\vspace{.3 cm}
\unitlength=1cm
\begin{picture}(5,0)
\put(-1.00,15.40){a)}
\put( 4.95,14.73){$\bf e_{1}$}
\put( 4.95,13.92){$\bf e_{2}$}
\put( 4.95,10.73){$\bf s_{1}$}
\put( 4.95, 9.98){$\bf s_{2}$}
\put( 1.50, 8.50){$\bfM$}
\put( 1.32,11.15){$\bf 0$}
\put( 1.32,10.43){$\bf 0$}
\put( 1.32, 9.50){$\bf 0$}
\put( 2.12,11.15){$\bf 0$}
\put( 2.12,10.03){$\bf 0$}
\put( 4.95,15.15){$\bf 0$}
\put( 4.95,14.22){$\bf 0$}
\put( 4.95,13.50){$\bf 0$}
\put( 4.95,11.20){$\bf 0$}
\put( 4.95,10.25){$\bf 0$}
\put( 4.95, 9.50){$\bf 0$}

\put(-1.00,8.20){b)}
\put( 1.55,3.62){$\bfUhit$}
\put( 2.85,3.62){$\bfMhit$}
\put( 1.30,6.60){$\bfWhit$}
\put( 2.80,8.00){$\bftVhit$}

\put(-1.00,2.80){c)}
\put( 2.35,2.03){$\bfe_{1}$}
\put( 2.35,1.57){$\bfe_{2}$}
\put( 0.95,0.20){$\bftVhit$}
\put( 2.70,0.80){$\bf =$}
\put( 5.10,2.30){$\bfs_{1}$}
\put( 5.10,1.70){$\bfs_{2}$}
\put( 3.70,0.20){$\bfRhit$}
\end{picture}
\caption{Inverse protocol for sub sets of the detector. a) Selection of the useful blocks of the initial matrix $\bfM$. The blocks labeled by a $\bf 0$ are composed of zeros only. b) Building of the $\bfMhit$ matrix from the selected blocks and SVD truncation. c) Reduced system to be solved ($\bfRhit={\bfWhit}^{-1}\,\bftUhit$).}
\label{Fig direct}
\end{center}
\end{figure}

A more rigorous procedure to solve the decomposition problem is first to select the useful blocks of the transform matrix then to calculate the singular value decomposition (fig. \ref{Fig direct}) for the selected blocks. In this case, the singular value decomposition has to be performed event-by-event, thus this methode is appropriate for off-line applications or when the acquisition rate is low. 

However, for a reduced number of segment combinations (the most probable ones), it is possible to pre-calculate off-line the $\bftVhit$ and $\bfRhit$ matrices resulting from the selection of the blocks followed by their singular value decomposition. This set of matrix pairs is saved in the computer memory so that it can be used on-line when the corresponding segment combination is encountered. For the combinations that were not memorized, the previous protocol is used. Table I shows that, due to computer memory limitation, this clustering method can be used for coaxial detectors only when each combination involves less than three or four adjacent hit segments.

\begin{table}[htbp]
\caption{ Number of possible segment combinations as a function of the number of hit segments in the case of 36 segment coaxial detectors. A combination gather segments which are close enough for their signals to overlap.}
\begin{center}
\begin{tabular}{|c|r|}
\hline
number of& number of \quad \\
hit segments   & combinations \\
\hline
 1 &  36 \\
\hline
 2 &  315\\
\hline
 3 &  3 146\\
\hline
 4 &  22 951\\
\hline
 5 &  137 957 \\
\hline 
\end{tabular}
\end{center}
\label{default}
\end{table}

%
%
\bibliographystyle{unsrt}

\end{document}